\documentstyle[12pt]{report}
\newtheorem{teorem}{Theorem}[chapter]
\newtheorem{proposisjon}[teorem]{Proposition}

\newtheorem{lemma}[teorem]{Lemma}
\newtheorem{korollar}[teorem]{Corollary}

\newenvironment{Proof}{{\em Proof:} \par}%
{\hfill\framebox[1em]{\rule[-.3ex]{0em}{1.0ex}\vspace{3ex}}\par}

{\hfill\framebox[1em]{\rule[-.3ex]{0em}{1.0ex}\vspace{3ex}}\par}

\makeatletter
\newcommand{\verbinput}[1]{{\@verbatim\frenchspacing\@vobeyspaces
	    \input{#1}\endtrivlist}}
\def\abstract{\if@twocolumn
\section*{Abstract}
\else \small
\begin{center}
{\bf Abstract\vspace{-.5em}\vspace{0pt}}
\end{center}
\quotation
\fi}
\def\endabstract{\if@twocolumn\else\endquotation\fi}
\makeatother
\newcommand{\beq}{\begin{eqnarray}}
\newcommand{\eeq}{\end{eqnarray}}
\newcommand{\beqq}{\begin{eqnarray*}}
\newcommand{\eeqq}{\end{eqnarray*}}

\newcommand{\bm}[1]{\mbox{\boldmath $#1$}}
\begin{document}
\phantom{jurek}
\vspace{2cm}
\def\thepage{}
\begin{center}
{\Large{\bf Thermal fluctuations around
\newline
classical crystalline ground states
\newline
The case of bounded fluctuations}}\\[2 mm]
\end{center}
\begin{center}
\large \bf
{Sergio Albeverio\footnote{
Fakult\"{a}t f\"{u}r Mathematik, Ruhr-Universit\"{a}t Bochum,
D-4630 Bochum 1, Germany}}\\[2 mm]
{Roman Gielerak\footnote{
Institute of Theoretical Physics, University of Wroclaw, 50-205 Wroclaw,
Poland.}
%BiBoS Research Center, University of Bielefeld, Bielefeld, Germany.}
}\\[2 mm]
{Helge Holden\footnote{
Institutt for matematiske fag, Norges Tekniske H\o gskole, Universitetet i
Trondheim,\\
\phantom{N-70}N-7034 Trondheim, Norway.}}\\[2 mm]
{Torbj\o rn Kolsrud\footnote{
Department of Mathematics, Royal Institute of Technology, S-10044 Stockholm,
Sweden.}}\\[2 mm]
{Mohammed Mebkhout\footnote{
Universit\'{e} d`Aix Marseille II, Facult\'{e} des Sciences de Luminy, F-13288
Marseille, France.}}\\
\end{center}
\begin{abstract}
We  study low--temperature  non Gaussian  thermal fluctuations	of a system of
classical particles  around a (hypothetical)  crystalline ground state.  These
thermal fluctuations are described by the  behaviour of a system of long range
interacting charged  dipoles at high--temperature  and high--density. For  the
case of  uniformly bounded fluctuations,  the low--temperature linked  cluster
expansion  describing  the  contribution  to  the  free  energy is derived and
analysed.   Finally  some   nonpertubative  results   on  the	existence  and
independence  of boundary  conditions of  the Gibbs  states for the associated
dipole systems are obtained.\\

\begin{description}
\item[Key words:]
crystalline ground states, low temperature expansion, long range dipole
systems,
thermal fluctuations.
\end{description}
\end{abstract}

%-----\\
%$^*$ Permanent address: Institute of Theoretical Physics, University of
%%Wroclaw, 50-205 Wroclaw, Poland.\\
\newpage
\def\thepage{\arabic{page}}

\chapter{Introduction}
The question why and how crystals form at low temperature belong
to the major open problems in theoretical physics [1,2,18].

Considering a system of $N$ classical particles described by some Hamiltonian
function $H$, we would like to know that the ground state configurations,
which correspond to the local minima of $H$, are very regular and that
under a small perturbation obtained by coupling this $N$--particle system
to a heat bath at temperature $T_r$, the typical particle
configurations for small values of $T_r$ do
not fluctuate much away from the crystallic ground state configuration.
It is clear that certain surface effects could destroy this naive
picture and therefore one has to pass to the thermodynamic
limit which corresponds to letting $N \uparrow \infty$ while keeping
the density of particles fixed. The
problem of verifying the validity of the above picture is still unsolved
in general, and in the situation
where the system is located in three dimensional space,
which is most interesting from a physical point of view,
there are very few results. However some progress has been made
in lower dimensions both for finite systems $(N < \infty)$ and in the
thermodynamic limit $(N \uparrow \infty)$.
The case of one space dimension seems to be fairly well
understood [3,4,6,7,20], but see [36]. For the case of a two-dimensional
space our understanding is more restricted [8,9,10] and in higher
dimensions only very few results are known [5,24,]. In one and
two dimensions also models with quasicrystalline behaviour at zero
temperature [11,12] have been studied. Quite recently a quantum
mechanical model showing crystallization has been discussed [13,14].
For a very readable survey of current insight into the crystal
problem we refer the reader to [15].

The present contribution is devoted to the analysis of the crystal
problem in three dimensional space. We select a special class of
interactions for which the analysis of the thermodynamical limit
of the free energy density describing fluctuations of a system at low
temperature around the hypothetical crystalline configuration, which
minimizes the relative potential energy, is possible by relatively
simple and standard methods. As we will see, the condition
for the minimizing potential energy configurations can be
expressed mathematically as a condition of positive definiteness
of the Hessian matrix for the corresponding potential function.
This enables us to rewrite the higher order contributions
of the Taylor expansion around a crystalline,
energy minimizing configuration in terms of certain Gaussian
integrals. In this picture, non Gaussian fluctuations around the
crystalline structure can be described by a highly nonlinear
interacting system of dipoles in which the potential describing
the interaction between the dipoles in the dipole gas framework
decays rather slowly, but the interaction energy is integrable.
A similarity with the corresponding problems of lattice systems
may be seen from this picture.

The continuous and the lattice dipole systems with long range like
interactions have been studied intensively only recently and a
variety of deep results describing their high and low temperature
properties have been derived rigorously.

We expect that some of the results obtained in [16,17,18,19,20]
can be applied to the dipole systems which we put in evidence in
our study of the formation of crystals in the three-dimensional space.

Restricting further the class of
admissible potentials we consider a class of potentials which
leads to bounded Taylor
remainders for the corresponding expansions around crystalline
ground state. For such	potentials the
associated dipole system describing fluctuations can be
interpreted as a grand canonical ensemble gas of
charged dipoles living on a Bravais lattice
$\Omega_\infty$ and in thermal equilibrium at
temperature $T_r^{-1}$ where $T_r$ is the temperature of
the real system. Thus the low temperature behaviour of a
real system is described
by the high temperature behaviour of the corresponding charged
dipole system. The activity of the associated dipole system is given by
$-\frac{1}{kT_{r}}$ where $k$ is the Boltzmann constant and
thus the low temperature behaviour of the real system is
described by the high temperature/high density limit of the
corresponding dipole system.
Section 3 of the present paper is devoted to the study of the
thermodynamic limit for the corresponding dipole systems using
some basic tools of statistical mechanics, e.g. linked cluster
expansion and Kirkwood--Salsburg identities [22,23]. Although
most of the results established there are valid for an ensemble
of dipoles of arbitrary sizes we shall restrict
ourselves to the study of restricted ensemble consisting of
dipoles of a uniformly bounded size. This cutoff is necessary for
obtaining a finite contribution to the free energy density of
the system we consider. The restriction to an ensemble of uniformly
bounded dipole is equivalent to the restriction of an
admissible size of possible deviations of real particles
configurations from cristalline configurations. From the
Gibbs measure point of view this means that we have restricted
ourselves to the contribution to the free energy density coming
from locally finite configurations with uniformly bounded (at
large distances) deviations from the (hypothetical) crystalline
ground state. The problem of large deviations which we called
large fluctuation problem is the main topic of a forthcoming,
second part of this work [34]. An additional discussion of the
meaning of the restriction to the bounded dipole case is included
in Section 4, where also a discussion of the
compatibility of our assumptions on the interaction
potential   paper can be found.

\chapter{Expansion around a crystalline ground state.
The finite volume case.}
Let us consider a system composed of $n$ classical particles
with masses equal to 1 (for simplicity) and enclosed in the
bounded region $\Lambda_N = \{x \in {\bm R}\,^3 \mid \,\, |x^i| \leq N$
for $i = 1,2,3\}$. The phase space coordinates of the system
will be denoted by
$(p, x)_{n} \equiv ((p_1, x_1),..., (p_n,x_n)),\,\,p_i \in
{\bm R}\,^3,\,\,x_i \in \Lambda_N$.
The particles interact through two-body potential forces
described by the potentials $V$. The class of
potential for which our analysis applies will be determined
by imposing certain conditions which will be listed
below. A preliminary discussion of the compatibility of these
hypotheses is given in section 4 of the present paper.\\[5 mm]

{\bf Hypothesis 0}\\
The potential $V$ is central, i.e. there exists a real-valued
function $\Phi$ on ${\bm R}\,^1_{+}$ such
that $V(x) = \Phi (\mid x \mid).$ The function $\Phi$ has
compact support and is of class at least $C\,\,^3$.
The potential $V$ is stable in the thermodynamical sense, i.e.,
\beq
\exists B>0, \forall n:\,\,\,{\cal E}((x)_n)
\equiv \sum_{1 \leq i < j\leq n} V(x_i - x_j) \geq -nB.
\eeq
The canonical (Maxwell-Boltzmann ensemble) partition
function of the system at (inverse) temperature $\beta > 0$ reads:
\beq
Z_N^n(\beta) = \int_{(R\,^{3}\times\Lambda_{N})^{\otimes n}}
\prod_{i=1}^n dp_i dx_i\,\exp [- \beta{\cal E} ((p,x)_n)],
\eeq
where the energy function is
\beq
{\cal E}((p,x)_n) \equiv \sum_{i=1}^{n} \frac{p_{i}^{2}}{2}
+ \sum_{i \neq j = 1}^{n} V(x_i - x_j).
\eeq
Integrating out the momenta we obtain:
\beq
Z_{N}^{n}(\beta) =
(\frac{2\pi}{\beta})^{3n/2}\int_{\Lambda_{N}^{\otimes n}}
\otimes_{i=1}^{n}dx_i\,\exp [- \beta {\cal E}((x)_n)].
\eeq
It is natural to try to apply the saddle point method to the
study of the low temperature $(\beta \uparrow \infty)$ behaviour
of the canonical partition function $Z_N^n(\beta)$. The
classical Laplace theorem tells us that if the minima of the
function ${\cal E}((x)_n)$ are well separated then the
following asymptotic formula is valid:
\beq
Z_N^n(\beta)_{\beta \uparrow \infty} \sim \, \sum_{X^{\min}_{n}}\,
e^{-\beta {\cal E}(X^{\min}_{n})}
\int_{\Lambda_{N}^{\otimes n}} \prod_{i=1}^{n}\,
\exp [-\beta \frac{1}{2} ((x_n) -
X^{\min}_{n}),[D^2{\cal E}]((x)_{n} - X_{n}^{\min}))] \cdot
\eeq
\beqq
\cdot \exp [- \beta {\cal R}(X_{n}^{\min} ; (x)_n)]\otimes^{n}_{i=1} dx_i.
\eeqq
where we have denoted by $X_{n}^{\min} = (x_{1}^{m},...,x_n^m)$
the minimizing configuration for the energy
function ${\cal E}$, and the contribution coming from the first
factor i.e. from $e^{- \beta {\cal E} (X_{n}^{\min})}$, dominates as
$\beta \uparrow \infty$.

The zero temperature crystal problem in this picture amounts to
answering the question whether the energy function
${\cal E}$ has a minimum at an (almost) regular configuration
$X_{n}^{\min} = (x_1^m,...,x_n^m)$. In the
finite volume situation we do not expect that the
minimizing configuration $X_{n}^{\min}$ forms a completely
regular structure, because of surface effects. To avoid such
problems we therefore pass to the thermodynamic limit, i.e.,
we consider the limit $N \uparrow \infty$, keeping
the density $\rho = n/(2N)^3$ fixed. It is expected that there
exists a certain regime of values and $\rho$ in which the
limiting minimizing configurations
$X_{\infty}^{\min} = (x^m_1,...,x_\infty^m)$
are quite regular globally. A particularly interesting
situation is the one where the (hypothetical) energy density
minimizing configuration $X^{\min}_{\infty}$ forms a
Bravais lattice in ${\bm R}\,^{3}$. \ \footnote{To deal with
the case $N = \infty$ it is convenient to use the energy density
function
\[\epsilon_{\rho}((x)_{\infty}) \equiv \lim_{\begin{array}{c}
N \uparrow \infty \\
\rho - fixed \end{array}}{\cal E}((x)_{n})  \frac{1}{(2N)^3}\]
provided it exists.}

Let us recall that a Bravais lattice
$\Omega_\infty$ in ${\bm R}\,^3$ is defined by three linearly
independent vectors
${\vec a}_1,\,{\vec a}_2,\, {\vec a}_3 \in {\bm R}\,^{3}$ such that
\beqq
\Omega_\infty \equiv \{ n_1{\vec a}_1 + n_2 {\vec a}_2
+ n_3{\vec a}_3 \mid n_1,\,n_2,\,n_3 \in {\bm Z}\}
\eeqq
Let us set $\Omega_N = \Lambda_N \cap \Omega_\infty$.
Assuming $n = \mid \!\Omega_N\!\mid$ we can, by
suitably relabeling and changing the variables, rewrite
the energy function of n-particles in the following
way:
\beq
{\cal E}((x)_n) = \frac{1}{2}\sum_{\begin{array}{c}
\lambda, \mu \in \Omega_{N}\\
\lambda \neq \mu \end{array}} V (\lambda - \mu + y_\lambda - y_\mu)
\eeq
where the variables $y_\lambda$ measure the deviation of
the $\lambda$ particle from the lattice point
$\lambda \in \Omega_N$. This can be further represented as:
\beq
{\cal E}((x)_n) = {\cal E}^{cr}(\Omega_N) + {\cal E}^{dev}
(\Omega_N \mid \{y_\lambda\}),
\eeq
where
\beq
{\cal E}^{cr}(\Omega_N) = \frac{1}{2} \sum_{\begin{array}{c}
\lambda \neq \mu\\
\lambda, \mu \in \Omega_{N} \end{array}} V(\lambda - \mu)
\eeq
is the energy of the crystalline configuration $\Omega_{N}$ and
\beq
{\cal E}^{dev}(\Omega_N \mid \{y_\lambda\}) = {\cal E}((x)_n)
- {\cal E}^{cr}(\Omega_N) \nonumber \\
= \frac{1}{2} \sum_{\begin{array}{c}
\lambda, \mu \in \Omega_{N}\\
\lambda \neq \mu  \end{array}} [V(\lambda - \mu + y_\lambda
- y_\mu ) - V(\lambda - \mu)]
\eeq
measures the energy deviation of the real configuration
$(x)_n$ from the energy of the crystalline configuration
$\Omega_{N}$.

The crystalline energy ${\cal E}^{cr}$ depends only on the parameters
$\vec{a}_1,\,\vec{a}_2,\,\vec{a}_3$ spanning our Bravais
lattice. Therefore one can easily find an equation on $\vec{a}_1,\,
\vec{a}_2\,\vec{a}_3$ that yields a stationary point for
the crystalline energy
${\cal E}^{cr}(\Omega_N)$ (see e.g., [21]). As an example,
in the case of central
potential forces described by the function $\Phi$ as in Hypothesis 0
the corresponding equation reads:
\beq
\sum_{\begin{array}{c}
n \in {\bm Z}\,^{3}\\
n \neq 0 \end{array}} \frac{n^{i}n^{j}\Phi '
(\mid \lambda \mid)}{\mid \lambda \mid} = 0,\,\,\,i,j = 1,2,3,
\eeq
which is a set of 9 equations with 9 unknowns.
This equation, together with the condition that
the relevant Hessian is positive, then gives
a local minimum for ${\cal E}^{cr}(\Omega_n)$.

We find that for technical reasons it is
more convenient to work with a periodic version of  the given finite
Bravais lattice, henceforth we shall understand under $\Omega_N$ this
periodic version.

The following hypothesis plays a crucial role in the present
attempt to understand the
formation of crystals in the low-temperature/high density
region in three dimensional space.\\[5 mm]

{\bf Hypothesis 1}\\
(H1$\alpha)$. There exists a number $\rho_{cr} > 0$ such that
for all $\rho \geq \rho_{cr}$ there exists a Bravais lattice solutions
$\Omega_{N}$ of the minimizing problem for the corresponding
${\cal E}^{cr}(\Omega_N)$. Moreover this holds
uniformly in $N$ at least for sufficiently large $N$. \\
(H1$\beta$). The deviation energy function
${\cal E}^{dev}(\Omega_N \mid \{y_n\}_{\lambda \in \Omega_{N}})$
has a stationary point at $\{y_\lambda = 0, \lambda \in \Omega_N\}$
which is a local minimum for sufficiently large values of $N$.

Part $\beta$ of Hypothesis 1 can be derived from the condition.
\beq
0 < (y,\,[D^2 V] y) \equiv \frac{1}{2}
\sum_{\lambda, \mu \in \Omega_{N}}(y_\lambda - y_\mu , [D^2 V]
(\lambda - \mu) (y_\lambda - y_\mu))
\eeq
where $[D^2 V] (\lambda - \mu)$ is the Hessian of $V$ evaluated
at the lattice point $(\lambda - \mu)$, see e.g., [21].

Condition (2.11) can be used to deduce the existence of a
certain functional integral
representation of the quantities of interest. For this
reason we define the matrix:
\beq
A_N (\lambda, \mu) = \left\{ \begin{array}{ll}
2 \sum_{\eta \in \Omega_{N},\,\lambda \neq \eta}\,[D^2 V]
(\lambda - \eta) & \mbox{ if }\,\,\,\lambda = \mu\\
- [D^2 V] (\lambda - \mu) & \mbox{ if }\,\,\,\lambda
\neq \mu. \end{array} \right.
\eeq
Then we have the following equality:
\beq
\frac{1}{2} \sum_{\lambda, \mu \in \Omega_{N}}
(y_\lambda - y_\mu\,, [D^2 V] (\lambda - \mu) (y_\lambda - y_\mu))
= \sum_{\lambda, \mu \in \Omega_{N}}(y_\lambda, A_N(\lambda, \mu)y_\mu)
\eeq
{}From the hypothesis H1$\beta$ it follows that the matrix
$A_N$ is strictly positive definite on the space
${\bf R}^{3\mid \Omega_{N}\mid}$ (with $\mid\!\Omega_N\!\mid$
the number of points in $\Omega_N)$. On the Borel $\sigma$-algebra
of sets of ${\bm R}\,^{3\mid\Omega_{N}\mid}$
we can therefore define the Gaussian measure $\mu_N^0$ with
mean equal to zero and covariance matrix
${\bm C}_N(\lambda,\mu)$ equal to the inverse of the matrix
$A_N(\lambda, \mu)$, i.e.,
\beq
\int_{{\bm R}\,^{3\mid\Omega_{N} \mid}}d\mu_N^0 (y_{1,...,}
y_{\mid\Omega_{N}\mid})
\exp [i \sum_{\lambda \in \Omega_{N}}(\alpha_\lambda, y_\lambda)]
\equiv {\bm E}_N^0 (e^{i(\tilde \alpha, \tilde y)})
\\ \nonumber
= \exp [- \frac{1}{2} (\alpha, A_N^{-1}\alpha)] =
\exp [- \frac{1}{2}(\alpha, {\bm C}_N\alpha)]
\eeq
where
\beq
\tilde{\alpha} = (\alpha_1,...,\alpha_{\mid\Omega_{N}\mid}),
\eeq
\beq
(\tilde{\alpha}, \tilde{y}) = \sum_{\lambda \in \Omega_{N}}
(\alpha_\lambda, y_\lambda),
\eeq
\beq
(\tilde{\alpha}, C_N \tilde{\alpha}) =
\sum_{\lambda \in \Omega_{N}}(\alpha_\lambda,
\sum_{\mu \in \Omega_{N}}
{\bm C}_N(\lambda, \mu)\alpha_\mu),
\eeq
and similarly for $(\tilde{\alpha}, A_N^{-1}\tilde{\alpha})$.
{}From the periodic lattice $\Omega_N$ and the infinitely
extended lattice $\Omega_\infty$ we form the corresponding
dual lattices $\Gamma_N$ and $\Gamma_\infty$ respectively.
The corresponding dual groups or
Brillouin zones are denoted by $\hat{\Omega}_N$ and
$\hat{\Omega}_\infty$ respectively and
are defined by
$\hat{\Omega}_N = [-N,N]^3/\Gamma_N,\,\,\,\hat{\Omega}_\infty
= {\bm R}\,^3/\Gamma_\infty$. We have:
\beq
\frac{1}{2}\sum_{\begin{array}{c}
\lambda,\mu \in \Omega_{N}\\
\lambda \neq \mu \end{array}} (y_\lambda - y_\mu, [D^2 V]
(\lambda - \mu)(y_\lambda - y_\mu))
\newline
= \sum_{p \in \hat{\Omega}_{N}} \overline{\hat{y}_{N}(p)}
(\hat{A}_N(0) - \hat{A}_N (p))\hat{y}_N(p),
\eeq
where
\beq
\hat{A}_N(p) \equiv \sum_{\lambda \in \Omega_{N},
\lambda \neq 0}[D^2 V] (\lambda) e^{ip \lambda}
\eeq
and
\beq
\hat{y}_{N}(p) \equiv \sum_{\lambda \in \Omega_{N}}
y_\lambda e^{ip\lambda}.
\eeq
For the infinitely extended lattice $\Omega_\infty$ the
corresponding formulae are:
\beq
\frac{1}{2} \sum_{\begin{array}{c}
\lambda,\mu \in \Omega_{\infty}\\
\lambda \neq \mu \end{array}}
(y_\lambda - y_\mu,\,[D^2 V] (\lambda - \mu)(y_\lambda - y_\mu))
\newline
= \int_{\hat{\Omega}_{\infty}} \overline{\hat{y}(p)}
(\hat{A}_\infty(0) - \hat{A}_\infty(p))\hat{y}(p)dp
\eeq
where
\beq
\hat{y}(p) = \sum_{\lambda \in \Omega_{\infty}}y_{\lambda}e^{ip\lambda}
\eeq
\beq
\hat{A}_\infty (p) = \sum_{\begin{array}{c}
\lambda \in \Omega_{\infty}\\
\lambda \neq 0 \end{array}} [D^2 V] (\lambda) e^{ip\lambda}
\eeq
We remark that, due to the compact support of $D^2V(\lambda)$ and
the fact that $\Omega_N$ is compact, $\hat{A}_\infty (p)$ is
well-defined, and is the pointwise limit of $\hat{A}_N(p)$
as $N \rightarrow \infty$.
Similarly, the right hand side of (2.22) is well-defined for
all $y \in L_2 (\Omega_\infty)$. Hence the
equalities (2.21), established first for a dense subset of
$L_2(\Omega_\infty)$, extend to
the whole of $L_2(\Omega_\infty)$.

Let us define the functional ${\cal E}^{dev}_{\Omega_{\infty}}$
on the dense subset of $L_2(\Omega_\infty)$ consisting of
sequences $\{y_\lambda\},\,\lambda \in \Omega_\infty$ with
compact support, by the formula (2.9) with
$N = \infty$, i.e., by
\beq
{\cal E}^{fl}(\Omega_{\infty} \mid
{\{y_\lambda\}}_{\lambda \in \Omega_\infty}) \equiv
\frac{1}{2} \sum_{\begin{array}{c}
\lambda, \mu \in \Omega_{\infty} \\
\lambda \neq \mu \end{array}}
[V (\lambda - \mu + y_\lambda - y_\mu ) -
V(\lambda - \mu)].
\eeq
Then we have, using formula (2.20), the following:

\begin{lemma}
Assume that the point
$\{y_\lambda = 0\},\,\,\,\lambda \in \Omega_\infty$ is a stationary
point of ${\cal E}^{dev}_{\Omega_{\infty}}$. Then it is a local
minimum for ${\cal E}^{dev}_{\Omega_{\infty}}$ if
$\hat{A}_\infty (0) - \hat{A}_\infty (p)$ is a positive definite
matrix for all $p \in \hat{\Omega}_\infty$. The corresponding
statements hold for ${\cal E}^{dev}_{\Omega_{N}}$ i.e., if
$\{y_\lambda\},\,\lambda \in \Omega_N$ is a
stationary point for ${\cal E}^{fl}_{\Omega_{N}}$ then it is a
local minimum if $\hat{A}_N (0) - \hat{A}_N (p)$ is a positive definite
matrix for all $p \in \hat{\Omega}_N$.
\end{lemma}
By Fourier analysis, similarly as above, we find the
following representation for the inverse matrices
$(A_N(\lambda, \mu))^{-1}$ and $A_\infty (\lambda, \mu)$.
\beq
{\bm C}_{N} (\lambda - \mu) \equiv (A_{N})(\lambda, \mu)
= \sum_{p\in \hat{\Omega}_N} e^{ip(\lambda - \mu)}
{(\hat{A}_N(0) - \hat{A}_N(p))^{-1}}
\eeq
and
\beq
{\bm C}_\infty (\lambda - \mu) \equiv (A_\infty^{-1})(\lambda, \mu) =
\int_{\hat{\Omega}_{\infty}}e^{ip(\lambda - \mu)}(\hat{A}_\infty (0)
- \hat{A}_\infty (p))^{-1}dp.
\eeq
The following hypothesis will play an essential role in the
following analysis.\\[5 mm]

{\bf Hypothesis 2}\\
(H2$\alpha)$. There exists a constant $c$, independent of $N$,
such that ${\hat A}_N(p) - \hat{A}_N(0) \geq
cp^2 + 0_N(p^2)$ as $p \rightarrow 0$.\\
(H2$\beta)$. ${\hat A}_N (p) = \hat{A}_N(0)$ iff $p = 0$.

Hypothesis 2 ensures that the functions ${\bm C}_N(\lambda - \mu)$ and
${\bm C}_\infty(\lambda - \mu)$ are well-defined and moreover
${\bm C}_\infty(\lambda - \mu)$ decays at least as
$\frac{1}{\mid \lambda - \mu \mid}$
for $\mid\!\lambda - \mu\!\mid \rightarrow \infty$\,.
The set of potentials for which (H2$\alpha)$ and
(H2$\beta)$ are satisfied is surprisingly rich, we refer to the
work [21] and to section 4 of the present
paper.

The previously introduced Gaussian measure $d\mu_N^0$ has
the following density
with respect to the Lebesgue measure
\beq
\frac{d\mu_N^0}{\mathrel{\mathop{\otimes}
\limits_{\lambda \in \Omega_N}}d y_\lambda} = \det [(\frac{1}{2\pi}
A_N(\lambda, \mu))_{\lambda, \mu \in \Omega_{N}}]^{-\frac{1}{2}}
\exp [-\frac{1}{2} \sum_{\lambda, \mu \in \Omega_{N}}
(y_\lambda, A_N(\lambda, \mu) y_\mu)]
\eeq
Expanding around the minimizing configuration $\Omega_N$, using
the Gaussian measure integration
${\cal E}_N^0$ and making a simple change of variables, we can
represent the canonical partition
function $Z_N^n$ in the following way:
\beq
Z_N^n(\beta) = (2\pi)^{3n/2} \beta^{-3n} \cdot Z^{cr}(\Omega_N)
\cdot z_N \cdot Z_N^{fl}(\beta),
\eeq
where we have introduced\\
\begin{itemize}
\item the crystalline partition function:
\beq
Z_\beta^{cr}(\Omega_N) \equiv \exp [- [\beta {\cal E}^{cr}(\Omega_N)]
\eeq
\item the fluctuation partition function:
\beq
Z_N^{fl} (\beta) \equiv {\bm E}_N^0
(\chi(\{\mid y_\lambda \mid \leq N\}_{\lambda \in \Omega_N}))
\exp [- \beta {\cal R}(\Omega_N \mid
\{\frac{y_{\lambda}}{\sqrt{\beta}}\}_{\lambda \in \Omega_{N}})]
\eeq
where
\beq
{\cal R}(\Omega_N \mid \{y_\lambda\}_{\lambda \in \Omega_{N}}) \equiv
{\cal E}((x)_n
\\ \nonumber
- \frac{1}{2} \sum_{\lambda, \mu \in \Omega_N}
(y_\lambda - y_\mu , D^2 V (\lambda - \mu) (y_\lambda - y_\mu))
- {\cal E}^{cr} (\Omega_N)
\eeq
is the Taylor remainder of the Taylor expansion of the energy
${\cal E}((x)_n)$ of the real configuration
$(x)_n$ around the local minimum given by $\Omega_N$
\item the normalisation factor for the measure $\mu_N^{0}$:
\beq
z_N \equiv \det (\frac{1}{2\pi}(A_{N}(\lambda - \mu)_{\lambda,
\mu \in \Omega_N})^{1/2}
\eeq
\end{itemize}
The free energy density of a system enclosed within the region
$\Lambda_N$ (with periodic boundary conditions) is defined by:
\beq
\beta P_N (\beta) = \frac{1}{\mid \Lambda_{N}\mid} \ln Z_{N}^{n}(\beta)
\eeq
Substituting (2.28) into (2.33) we obtain:
\beq
\beta  P_N(\beta) = \frac{3 \ln 2\pi}{2}\rho - 3 \rho \ln \beta
+ \beta P_N^{cr}(\Omega_N \mid\beta) + \beta \tilde{z}_N
+ \beta p_N (\beta)
\eeq
where
\beq
P_N^{cr}(\Omega_N \mid \beta) = \frac{1}
{\beta \mid \Lambda_N \mid} \ln \exp [-\beta {\cal E}^{cr}(\Omega_N)]
= - \frac{{\cal E}^{cr}(\Omega_{N})}{\mid \Lambda_N \mid}
\eeq
is the crystalline free energy density;
\beq
\tilde{z}_N \equiv \frac{1}{2\beta} \frac{1}{\mid \Lambda_N\mid} \ln \det
((A_N \frac{(\lambda, \mu))}{2\pi} \lambda, \mu \in \Omega_N)
\eeq
is a numerical constant and
\beq
p_N(\beta) \equiv \frac{1}{\beta \mid \Lambda_N \mid} \ln {\bm E}_N^0
[\chi (\{\mid y_\lambda \mid \leq N\}_{\lambda\in \Omega_N)}
\exp [- \beta {\cal R}(\Omega_N \mid
\left\{ \frac{y_{\lambda}}{\sqrt{\beta}} \right\} )]]
\eeq
is the free energy density of nongaussian fluctuations around $\Omega_N$.

A simple Fourier analysis is used to control the thermodynamic
limit of the quantity $\tilde{z}_N$. We have
\beq
\tilde{z}_N = \frac{1}{2\beta} \frac{1}{\mid \Lambda_N \mid} \ln
\det ((A_N \frac{(\lambda, \mu))}{2\pi} \lambda, \mu \in \Omega_N)
\\ \nonumber
= \frac{1}{2\beta} \frac{1}{\mid \Lambda_N \mid}
\sum_{p \in \hat{\Omega}_{N}} \mbox{ tr } \ln
(\hat{A}_N(0) - \hat{A}_N(p)).
\eeq
As $N \rightarrow \infty$ \,this tends to:
\beq
z_\infty = \frac{1}{2\beta\mid \hat{\Lambda} \mid}
\int_{\hat{\Lambda}}dp \mbox{ tr } \ln [\hat{A}_\infty(0)
- \hat{A}_\infty(p)],
\eeq
which is finite, e.g., under the assumption (H2$\alpha$).

The most difficult element of our analysis concerns the problem
of the existence of the
thermodynamic limit of the free energy density of fluctuations
around a crystalline ground state. Taking
into account the formula
\beq
{\cal R}(\Omega_{N} \mid \{\frac{y_{\lambda}}
{\sqrt{\beta}} \}_{\lambda \in \Omega_{N}}) =
\sum_{\lambda, \mu \in \Omega_{N}}{\cal R}_{\Omega_{N}}
(\lambda - \mu \mid
\frac{y_\lambda - y_\mu}{\sqrt{\beta}}),
\eeq
where
\beq
{\cal R}_{\Omega_{N}}(\lambda - \mu \mid \frac{y_\lambda - y_\mu}
{\sqrt{\beta}})
\equiv V(\lambda + \beta^{-1/2}y_{\lambda} - \mu
- \beta^{-1/2}y_{\mu}) \\ \nonumber
- V (\lambda - \mu) - \frac{1}{2} \frac{(y_\lambda - y_\mu)}{\sqrt{\beta}} \,
[D^2V](\lambda - \mu) ( \frac{y_\lambda - y_\mu}{\sqrt{\beta}})),
\eeq
we see that in the field theoretical picture given by the Gaussian
integral representation (2.30) the partition function
$Z_N^{fl}$ can be viewed as
the partition function of a gas of dipoles that sit on the lattice
$\Omega_N$ and interact by a nonpolynomial and in general nonlocal
potential given by (2.40). With some
additional simplifying assumptions the problem of the thermodynamic
limit of the free energy density and also for the corresponding
Gibbs states describing the associated
dipole systems will be studied in the next section by an
application of techniques of statistical mechanics
e.g. linked cluster expansion and analysis of the
Kirkwood-Salzburg identities.

\chapter{Infinite volume limit. Bounded dipole length case.}
In this section we shall discuss the thermodynamic limits of the
free energy density and the Gibbs states of the associated dipole
systems describing bounded fluctuations of the real configurations around
the crystalline ground state given by a Bravais lattice
$\Omega_\infty$.

{\bf 3.1 A convergent expansion for the dipole partition function}
In order to simplify the analysis as much as
possible the following additional hypothesis will be introduced.

{\bf Hypothesis 3.}
There exists a family $(d\mu_\lambda)_{\lambda \in 0_{R}}$ of
complex measures on ${\bm R}\,^3$ such that the following
representation for the corresponding Taylor remainders defined by
(2.42) hold for all $\lambda - \mu $:
\beq
{\cal R}_{\Omega_{N}}(\lambda - \mu \mid x) = \int_{R\,^{3}}
d\mu_{\lambda - \mu}(\alpha) e^{i\alpha x}.
\eeq
In addition we assume that
$d \bar{\mu}_\lambda (\alpha) = d\mu_\lambda(-\alpha)$ so that
${\cal R}_{\Omega_{N}}(\lambda - \mu \mid x)$ is real. We also
make the following additional assumptions on the measure appearing in
(3-1)$^1$.

(H3)$\alpha)$.
$\int d \mid\!\mu_\lambda\!\mid (\alpha ) \mid \alpha\mid^n < \infty,$
for all $\lambda$ and all $n \in {\bm N}$\ \footnote{In fact
we need only the existence of the first moments
(see Proposition 3-2 below)} \\
(H3$\beta)$. $\int d\mid\!\mu_\lambda\!\mid (\alpha ) e^{A\alpha^{2}}$,
for some $A > 0$, and all $\lambda \in \Omega_\infty$. $A$ has
to be sufficiently large (see below).\\
(H3$\gamma)$. For all $\lambda$, $\mbox{ supp } d\mu_\lambda$
is compact in ${\bf R}\,^3$.

The following trivial chain of implications holds:
(H3$\gamma) \Rightarrow (H3\beta) \Rightarrow (H3\alpha)$.
Which one of the hypothesis $\alpha , \beta$ and $\gamma$ is needed
depends on the particular situation we investigate (see below).

As a preliminary step in our analysis we remove the characteristic
function restricting the possible values
of the size of fluctuations $\{y_\lambda\}$ in (2.30) defining
\beq
Z_N^{d} \equiv {\bm E}_N^0 (\exp [-\beta {\cal R}
(\Omega_N \mid \{\beta^{-1/2}y_\lambda \}_{\lambda \in \Omega_N}]).
\eeq
Using the formula
\beq
{\bm E}_N^0 (\prod_{j=1}^n (e^{i\beta^{-1/2}\alpha_{j}y_{\lambda_{j}}}
\cdot e^{-i \beta^{-1/2} \alpha_j y_{\mu_{j}}})
= \exp [-\frac{1}{\beta} \sum_{1\leq i, j \leq n} {\bm V}_N
(\lambda_i \mu_i \alpha_i \mid\lambda_j \mu_j \alpha_j)]
\eeq
where
\beq
{\bm V}_N (\lambda_i \mu_i \alpha_i \mid \lambda_j \mu_j \alpha_j)
\equiv \alpha_i ({\bm C}_N (\lambda_i - \lambda_j)
+ {\bm C}_N (\mu_i - \mu_j)
\\ \nonumber
- {\bm C}_N (\lambda_i - \mu_j) -
{\bm C}_N (\mu_{j}-\lambda_i))\alpha_j
\eeq
and taking into account H4 and H6 we obtain the following expansion
of $Z_N^d$:
\beq
Z_N^{d} = \sum_{n \geq 0} \frac{(-\beta)^{n}}{n!}
\sum_{\begin{array}{c}
\lambda_i, \mu_i \in \Omega_{N}\\ \mid \lambda_i - \mu_i \mid \leq R
\end{array}}
\int \otimes^n_{l=1} d\mu_{\lambda_{l}- \mu_{l}}(\alpha_i)
\\ \nonumber
\exp [-\frac{1}{\beta} \sum_{1 \leq i, j \leq n} {\bm V}_N
(\alpha_i \lambda_i \mu_i \mid \alpha_j \lambda_j \mu_j)]\,.
\eeq
{}From the very simple inequality
\beq
\mid {\bm E}_N^0 (\exp [i \sum_\lambda \alpha_\lambda y_\lambda ]
\mid \leq 1
\eeq
we deduce the following estimate on $Z_N^{d}$:
\beq
Z_N^{d} \leq \exp  [ \mid \beta\mid \mu^* \mid \tilde{D}_N(1) \mid]
\eeq
where
$$
\mu^* \equiv \sup_{\lambda \in \Omega_\infty} \{\int d\mid
\mu_\lambda \mid \},
$$
and $\mid \tilde{D}_N(1) \mid$ is the cardinality of the set
$\Omega_N \times \Omega_N$. Estimate (3.7) shows that the
expansion (3.6) is absolutely convergent for any $\beta > 0$.
For the dipole free energy density \\
$p_N^d \beta) \equiv \mid \tilde{D}_N(1) \mid ^{-1} \ln Z_N^{fl}$
we obtain the upper bound:
\beq
p_N^d (\beta) \leq \beta \mu^*
\eeq
uniformly in $N$.\\
Although some of the results below hold for a dipole systems
described by the dipole partition function (3-3) it follows from
the estimate (3.8) that in order to ensure the finiteness of the
limit
$$\lim \frac{1}{\mid \Lambda_N \mid } ln
Z^d_N
$$
(see for (2.37)): without further assumptions on the interaction
potential $V$ we have to restrict the admissible length of
dipoles. Let $R$ be some fixed positive number. Then we define a
bounded length dipole system by the following partition function
\beq
Z^{bd}_N (\beta) \equiv {\bm E}^0_N (\exp - \beta
R^\mu_N (\Omega_N \mid \{ \beta^{- \frac{1}{2}} y_\lambda \} )
\eeq
where the restricted Taylor remainder $R^\mu_N$ is given by:
\beq
R^\mu_N (\Omega_n \mid \{ \beta^{- \frac{1}{2}} y_\lambda )
\equiv \sum_{\begin{array}{c} \lambda_i, \mu_i \in \Omega_N\\
\mid \lambda_i - \mu_i \mid \leq R \end{array}}
R_N (\lambda - \mu \mid \frac{ \{y_\lambda -
y_\mu\}}{\sqrt{\beta}} )
\eeq
{}From now on an allowed maximal size of dipole given by a number
$R$ will be fixed.
The following sort of stability also comes from estimate (3.7) in a
simple way:
\beq
\sum_{1 \leq i \neq j \leq n} {\bm V}_N (\alpha_i \lambda_i
\mu_i \mid \lambda_j \mu_j \alpha_j )
\geq - 2\sum_{i=1}^n \alpha_i ({\bm C}_N(0) - {\bm C}_N
(\lambda_i - \mu_i)) \alpha_i .
\eeq
The expansion like (3.6) for $Z^{bd}_N$ has the interpretation
as a grand canonical partition function of a system
of dipoles of length bounded by $R$ on the lattice $\Omega_N$
in the thermal equilibrium
at inverse temperature $1/\beta$. Because of the crucial
stability estimate (3.11) the standard tools of classical
statistical mechanics, such as cluster expansions,
Kirkwood-Salzburg type analysis [22] can be used to study the
corresponding thermodynamical limit. In order to simplify our
notation the following abbreviations will
be used:
\beq
D_N^R (K) \equiv D_N^R(1)\times...\times D_N^R(K) \\ \nonumber
= \{ (\lambda_1, \mu_1, \alpha_1),...,(\lambda_K, \mu_K, \alpha_K)
\mid \lambda_i, \mu_i \in \Omega_N, \mid \lambda_i - \mu_i
\mid \leq R, \alpha_i \in R^3 \}
\eeq
$\equiv K$-dipoles configuration space.\\
Furthermore we denote by
\beq
\tilde{D}_N^R(K) = \tilde{D }_N^R(1) \times...
\times\tilde{D}_N^R(K) \,\,\,\mbox{(see (3.10))}
\eeq
the $K$-dipoles restricted configuration space.\\
For a generic $\omega \in D_N^R(K)$ the following abbreviations
will also be employed
\beq
\omega = ((\lambda_1, \mu_1, \alpha_1),...,
(\lambda_n, \mu_n, \alpha_n)) \equiv (\lambda, \mu, \alpha)_n,
D_N^R(1) \ni (\lambda_i, \mu_i, \alpha_i) \equiv d(i).
\eeq
\beq
\int d_N^R(i)...\equiv \sum_{\begin{array}{c}
\lambda_i, \mu_i \in \Omega_N\\
\mid \lambda_i - \mu_i \mid \leq R \end{array}}
\int d\mu_{\lambda_{i}- \mu_{i}}(\alpha_i)
\eeq
\beq
\int d_N^R(1,...,n)... = \int d_N^R(n)... \int d_N^R(1)...
\eeq
\beq
\int d_\infty^R(i)... = \sum_{\begin{array}{c}
\lambda_i , \mu_i \in \Omega_\infty \\
\mid \lambda_i - \mu_i \mid \leq R \end{array}}
\int d \mu_{\lambda_{i} - \mu_{i}}(\alpha_i)...
\eeq
\beq
{\cal E}_N((\alpha, \lambda, \mu )_n) \equiv
\sum_{1 \leq i, j \leq n} {\bm V}_N (\alpha_i, \lambda_i, \mu_i
\mid \alpha_j \lambda_j \mu_j )
\eeq
\beq
{\cal E}_\infty((\alpha, \lambda, \mu )_n) \equiv
\sum_{1 \leq i, j \leq n} {\bm V}_\infty (\alpha_i \lambda_i \mu_i
\mid \alpha_j \lambda_j \mu_j )
\eeq
\beq
{\cal E}_N((\alpha, \lambda, \mu )_n \mid
(\alpha', \lambda', \mu' )_m) \\ \nonumber
\equiv {\cal E}_N((\alpha, \lambda, \mu, )_n \cup
(\alpha', \lambda', \mu' )_m)
- {\cal E}_N((\alpha, \lambda, \mu )_n) - {\cal E}_N
((\alpha', \lambda', \mu' )_m)
\eeq
and similarly for ${\cal E}_\infty$.\\
Hence we can rewrite (3.6) for $Z^{bd}_N$ in the following way
\beq
Z_N^{bd} = \sum_{n \geq 0} \frac{(-\beta)^{n}}{n!}
\int d_N^R (1,...,n) \exp [- \frac{1}{\beta}
{\cal E}_N(d(1),...,d(n))]
\eeq
{\bf 3.2
The linked cluster expansion for the free energy density.}\\
Using a well known trick of Mayer [22,23] we can derive the linked
cluster expansion for the free
energy density for a finite volume system. For this purpose we write
\beq
\exp [- \frac{1}{\beta} {\cal E}_N( d(1),...,d (n)) =
\prod_{i=1}^n V_N (i) \prod_{1 \leq i \neq j \leq n}
\exp [- \frac{1}{\beta} {\bm V}_N(d(i)\mid d(j))]
\\ \nonumber
= \prod_{i=1} V_N(i) \prod_{1 \leq i \neq j \leq n}
[(\exp - \frac{1}{\beta} {\bm V}_N (d(i) \mid d(j))] - 1) + 1]
\\ \nonumber
= \prod_{i=1}^n V_N (i) \sum_{\Gamma \subset \{1,...,n\}}
{\cal M}(d(i \in \Gamma))
\eeq
where
\beq
{\cal M}(d(i \in \Gamma)) \equiv \prod_{i,j \in \Gamma}
(\exp [-\frac{1}{\beta} {\bm V}_N(d(i)\mid d(j)] - 1)
\eeq
and
\beq
V_N(i) \equiv \exp [- \frac{1}{\beta} \alpha_i^2(C_N(0)
- C_N(\lambda_i - \mu_i))]
\eeq
is the corresponding vertex function.\\
We denote by ${\cal G}_n^N$ the set of all connected linear
$n$-graphs that can be built on the set $\tilde{D}_N^R(n)$.
Putting the expansion (3.23) into (3.22) and resumming we obtain
the linked cluster expansion in the following standard form:
\beq
Z_N^{bd} = \exp [\sum_{n \geq 1} (-\beta)^n \cdot b_N^n]
\eeq
where
\beq
b_N^n = \sum_{\Gamma \in {\cal G}_{N}}b_N^n(\Gamma)
\eeq
and the contribution $b_N^n(\Gamma)$ from the graph
$\Gamma \in {\cal G}_n^N$ is given by
\beq
b_N^n(\Gamma) = \frac{1}{\mid\tilde{D}_{N}^{R}\mid}
\int d_N^R(1)...\int d_N^R(n) \prod_{i\in {\cal L}(\Gamma)}
[\exp - \frac{1}{\beta} {\bm V}_N(l) - 1]
\prod_{i\in V(\Gamma)}V_N(i)
\eeq
where we have denoted the set of lines of a given
$\Gamma \in {\cal G}^N_n$ by ${\cal L}(\Gamma)$, by
$V(\Gamma)$ the set of vertices and taking $l \in \Gamma$ we denote
by $l_0$ the initial and resp. by $l_e$ the
endpoint of $l$ and then we define
${\bm V}_N (l) \equiv {\bm V}_N(l_0 \mid l_e)$.\\
{}From the positive definiteness of ${\bm C}_N$ it follows that
${\bm C}_N(0) - {\bm C}_N(\mu) \geq 0$ for
any $\mu \in \Omega_N$ and uniformly in $N$. Therefore we can
estimate the vertex function contribution
to (3.26) by:
$\mid \prod_{i \in V(\Gamma)} V_N (i) \mid \leq 1.$
Our first result for the dipole free energy density in the
thermodynamic limit $N \rightarrow \infty$ is the following.
\begin{proposisjon}
Assume that the hypotheses H2, H3 and H3$\gamma$ are all valid.
Then:
\beq
\forall_n \forall \Gamma \in {\cal G}_n
\overline{\lim}_{N \uparrow \infty} \mid b_N^n (\Gamma) \mid
< \infty \mbox{ for any }\,\,\,\beta \in {\bm R}_+ - \{0\}.
\eeq
\end{proposisjon}
\begin{Proof}
Any connected, linear $n$-graph $\Gamma \subset {\cal G}_n$
contains at least one spanning tree st$(\Gamma)$. Let \\
$\mid {\cal L}(\Gamma)\mid = k + s$, where $k = n - 1$ is the
number of lines forming st$(\Gamma)$ and
$s = \mid {\cal L}(\Gamma) \mid -n+1$ is the number of lines in
$\Gamma \backslash st(\Gamma)$. The contribution coming from the line
$l \notin st(\Gamma)$ can be estimated by a "$\sup$ argument":
\beq
\sup_l \mid \exp [\frac{1}{\beta}{\bm V}_N(l)] - 1 \mid
\leq 4 \beta^{-1}{\bm C}\,^*
\mid \alpha_{l_{0}} - \alpha_{l_{e}} \mid
\exp [\beta^{-1} 4\alpha_{i_{0}}\alpha_{i_{e}} {\bm C}\,^*]
\eeq
where
\beq
{\bm C}\,^* \equiv \sup_N (\sup_\mu \parallel {\bm C}_N(\mu)
\parallel )
\eeq
Setting
\beq
\alpha^* \equiv \sup_{\lambda \in 0_R} \sup
\{ \mid\!\alpha\!\mid \mid \alpha \in \mbox{ supp } d
\mu_\lambda\}
\eeq
it follows
\beq
\mid b_n^N (\Gamma)\mid \leq E(\beta)^sb_n^N( \mid st \Gamma
\mid )
\eeq
where
\beq
E = 4\beta^{-1}{\bm C}^*\alpha^{*2}\exp [\beta^{-1}4\alpha^{*2}C^*]
\eeq
and
\beq
b_n^N (\mid st \Gamma \mid ) = \frac{1}{\mid \tilde{D}_n^N(1)\mid}
\int d_N^R\mid (1,...,n)
\mid \prod_{i=1}^{n-1}\mid (\exp [-\frac{1}{\beta} {\bm V}(d(i)
\mid d(i+1))] - 1)\mid\,.
\eeq
Proceeding further we obtain
\beq
\mid b_n^N(\mid st\Gamma \mid ) \mid \leq \frac{1}{\mid
\tilde{D}_{N}^{R}(1)\mid} E'^{(n-1)}
\int \mid d_N^R (1,...,n) \mid \prod_{i=1}^{n-1} \mid
{\bm V}_N(d(i)\mid d(i+1)\mid
\eeq
where now
\beq
E' = \exp [\frac{4}{\beta} \alpha^{* 2}{\bm C}^*].
\eeq
Let $T_\rho$ be a translation by $\rho \in \Omega_N$. Observing
that the equality
$T_\rho{\bm V}_N(\lambda\mu \alpha \mid \lambda' \mu' \alpha')
\equiv {\bm V}_N(\lambda + \rho , \mu + \rho , \alpha \mid \lambda'
+ \rho , \mu' +\rho , \alpha') =
{\bm V}_N (\lambda \mu \alpha \mid \lambda \mu \alpha)$ holds, we
can estimate
\beq
\overline{\lim}_{N\uparrow \infty} \mid \tilde{D}_N^R(1)\mid^{-1}
\int \mid d_N^R (1,...,n) \mid \prod_{i=1}^{n-1} \mid {\bm V}_N(d(i)
\mid d(i+1)) \mid \\ \nonumber
\leq O(\beta) \sum_{\lambda_{1}\in 0_{R}}\sum_{\begin{array}{c}
\lambda_2,\mu_2 \in \Omega_{\infty}\\
\mid \lambda_2 - \mu_2 \mid \leq R \end{array}}
... \sum_{\begin{array}{c}
\lambda_n,\mu_n\\
\mid\lambda_n - \mu_n \mid \leq R \end{array}} \\ \nonumber
\int d \mid \mu_{\lambda_1 - \mu_1} \mid (\alpha_1) \cdot \int
d\mid \mu_{\lambda_{2}- \mu_{2}}\mid (\alpha_2)...
\int d\mid \mu_n - \mu_n \mid (\alpha_n)
\\ \nonumber
\mid {\bm V}_\infty (0, \lambda_1, \alpha_1 \mid\lambda_2 \mu_2
\alpha_2 ) ...
{\bm V}_\infty (\lambda_{n-1} \mu_{n-1} \alpha_{n-1} \mid
\lambda_n \mu_n \alpha_n) \mid
\\ \nonumber
\leq O'(\frac{1}{\beta}) \sum_{\lambda_{1} \in 0_{R}}
\sum_{\begin{array}{c} \mid \lambda_2 - \mu_2 \mid < R\\
\lambda_2,\mu_2\in \Omega_\infty \end{array}}\{\int d\mid
\mu_{\lambda_1}\mid (\alpha_1) d\mid
\mu_{\lambda_{2} - \mu_{2}} \mid (\alpha_2)
\\ \nonumber
{\bm V}_\infty (0, \lambda_1, \alpha_1 \mid \alpha_2, \mu_2,
\alpha_2) \}^{n-1}
\eeq
where $O(\frac{1}{\beta})$ and $O'(\frac{1}{\beta})$ are numerical
constants depending possibly of $R$
but not of $N$. We finish the proof if we show
\beq
\sum_{\lambda_1 \in 0_R} \sum_{\begin{array}{c}
\lambda_2, \mu_2 \in \Omega_\infty\\
\mid \lambda_2 -\mu_2 \mid \leq R. \end{array}}
\int d\mid \mu_{\lambda_{1}}\mid (\alpha_1) d \mid
\mu_{\lambda_{2}-\mu_{2}} \mid (\alpha_2) \mid \\ \nonumber
\cdot {\bm V}_\infty (0, \lambda_1, \alpha_1 \mid \lambda_2,
\mu_2,\alpha_2)\mid < \infty
\eeq
This amounts to study the decay properties of
$V_\infty(0,\lambda_1 \mid \lambda_2, \mu_2) =
{\bm C}_\infty(\lambda_2) - {\bm C}_\infty (\mu_2)
+ {\bm C}_\infty(\lambda_1 - \mu_2) - {\bm C}_\infty (\lambda_1
- \lambda_2)$ as $\mid \lambda_2 \mid \uparrow \infty$.\\
Let us define
\beq
\tilde{C}_\mu (\lambda) = {\bm C}_\infty(\lambda)
- {\bm C}_\infty(\lambda - \mu).
\eeq
Taking the Fourier transform of (3.40), using (2.21) and
that
${\bm C}_\infty(\lambda) = {\bm C}_\infty(-\lambda)$ we obtain
\beq
\hat{\tilde{C}}_\mu(p) = (1-\frac{1}{2} \cos p \mu)(A_\infty(0)
- A_\infty(p))^{-1}.
\eeq
{}From the hypothesis (H2$\beta$) it follows that
$\hat{\tilde{C}}_\mu(p)$ stays bounded as $p \rightarrow 0$
whenever $\mu \neq 0$. Therefore by the Fourier-Tauberian theorem
it follows that $\tilde{C}_\mu(\lambda)$ has to decay at least
as $\mid \lambda \mid^{-3}$ as $\mid \lambda \mid \uparrow \infty$.
Therefore taking $\mu ' \neq 0$, we conclude that the function
$\tilde{\tilde{C}}_{\infty,\mu}(\lambda) =
\tilde{{\bm C}}_\mu(\lambda) - \tilde{{\bm C}}_\mu(\lambda - \mu')$
has to decay at least as fast as $\mid \lambda \mid^{-4}$ as
$\mid \lambda \mid \uparrow \infty$ which is integrable over
${\bm R}\,^3$. Let us
apply this information to the problem (3.39). Using the definitions
of $\tilde{{\bm C}}_\infty$ and $\tilde{\tilde{{\bm C}}}_{\infty , \mu}$
\beq
\sum_{\mu_1 \in 0_N} \sum_{\begin{array}{c}
\mu_2, \lambda_2\in \Omega_\infty\\
\mid \lambda_2 - \mu_2 \mid \leq R \end{array}}
\mid {\bm V_\infty}(0 \mu_1 \alpha_1 \mid \lambda_2 \mu_2
\alpha_2 )\mid \\ \nonumber
\leq \mid \alpha_1 \mid \cdot \mid \alpha_2 \mid \cdot
\sum_{\mu_{1}} \sum_{\begin{array}{c}
\mu_2, \lambda_2 \in \Omega_\infty\\
\mid \lambda_2 - \mu_2 \mid \leq R \end{array}}
\mid \tilde{{\bm C}}_{\infty, \mu_{1}}(\lambda_2) -
\tilde{{\bm C}}_{\infty, \mu_1}(\mu_2) \mid \\ \nonumber
\leq \mid \alpha_1 \mid \cdot \mid \alpha_2 \mid \cdot O''(1) \cdot
\sum_{\lambda_2 \in \Omega_{\infty}} \mid
\tilde{\tilde{{\bm C}}}_{\infty , \mu}(\lambda_2) \mid < \infty
\eeq
for some $\mu \neq 0$.
\end{Proof}

{\bf Remark}\\
The above Prop. 3.1 can also be proven, essentially by the
same method, replacing assumption H3$\gamma$ by H3$\beta$.

Let us define the virial coefficients $b^n_\infty (\Gamma)$ and
the limit $N \uparrow \infty$ as the limit for $N \uparrow \infty$
of $b_N^n (\Gamma)$ as given by (3.28). Moreover by the
translational invariance we impose the restriction on $\Gamma$
that at least one of it vertex has to be located at some
$(0,\lambda)$ where $\lambda \in O_R$. That the limit
$\lim_{N \uparrow \infty}b_N^n (\Gamma)$ exists follows
from the simple observation that for a fixed $\Gamma$, the
sequence $(b_N^n(\Gamma))_N$ is a Cauchy sequence in N.

Let us remark that the limiting virial coefficient
\beqq
b^n_\infty (\Gamma) = \sum_{\begin{array}{c}
\mu_{1} \in O_{R} \\
(0,\mu_{1}) \in {\cal L}( \Gamma ) \end{array}}
\int d\mu_{\mu_1}(\alpha_1) \int d^R_\infty(2,\ldots ,n)
\prod_{l \in {\cal L}(\Gamma)}
[exp(- \frac{1}{\beta} V_\infty (l)-1] \times
\eeqq
\beq
\times \prod_{i \in V(\Gamma)} V (i)
\eeq
are analityc functions in $\frac{1}{\beta}$ for $\beta \neq 0$
and moreover $\lim_{\beta \rightarrow \infty}b^n_\infty (\Gamma)=0$.
{}From the Taylor expansion it follows that:
\beq
\frac{d^{M}b^n_\infty (\Gamma )}{d(-\frac{1}{\beta})^M}
= \sum_{\begin{array}{c}
k_1 , \ldots , k_{\mid {\cal L} (\Gamma) \mid} \\
k_i \geq 0 \\
\sum_{i=1}^{\mid {\cal L} (\Gamma)\mid } k_i +	\end{array}}
\sum_{\begin{array}{c}
r_{j}, \ldots, r_{\mid V(\Gamma) \mid}\\
r_{i} \geq 0 \\
\sum_{j=1}^{\mid V(\Gamma) \mid} r_j = M \end{array}}
\frac{(-\frac{1}{\beta})^{\sum_{i=1}^{\mid {\cal L}(\Gamma) \mid}
k_i + \sum_{j=1}^{\mid V(\Gamma) \mid} r_j -1}}
{\prod_{i=1}^{\mid {\cal L} (\Gamma)\mid} k_i !
\prod_{j=1}^{\mid V(\Gamma)\mid} r_j !}
\eeq
\beqq
\sum_{\begin{array}{c}
\mu_1 \in O_R \\
(0,\mu_{1}) \in {\cal L}(\Gamma) \end{array}}
\int d\mu_{\mu_1} (\alpha_1) \int d^R_\infty (2, \ldots ,n)
\prod_{l=1}^{\mid {\cal L} (\Gamma) \mid}
V_{\infty} (l)^{k_i} \prod_{j=1}^{\mid V(\Gamma) \mid}
V_{\infty} (j)^{r_j}
\eeqq
\beqq
\cdot \prod_{l \in {\cal L}(\Gamma)}[exp(- \frac{1}{\beta}
V_\infty (l)-(1-\theta (k_l))] \prod_{i \in V(\Gamma)} V(i)
\eeqq

from which it follows that also
\beq
\lim_{\beta \rightarrow \infty} \frac{d^M b^n_{\infty} (\Gamma)}
{d (- \frac{1}{\beta})^{M}} \equiv 0
\eeq
Since the total number of n-graphs $\Gamma$ restricted as
above is finite, we conclude
that the virial coefficients $b^n_\infty$ are analytic functions
of $\beta$ for $\beta \neq 0$ and moreover
$lim_{\beta \rightarrow \infty} b^n_\infty (\beta)=0$ and also
\beqq
\lim_{\beta \rightarrow \infty} \frac{d^M b^n_{\infty} (\beta)}
{d(-\frac{1}{\beta})^M} =0
\eeqq
 for any value of M.

A very convenient way to describe the virial coefficients in the
limit $N \rightarrow \infty$ has been given by Brydges and
Federbush [24].
We adopt their method to the case of the dipole systems to give the
linked cluster expansion for $p_\infty(\beta)$.
Towards this goal let us denote by $\Gamma_n$ the set of tree
functions $\eta$, defined on the set $\{1,...,n\}$, i.e.,
$\eta \in T_n \Leftrightarrow \eta: \{1,...,n\}
\rightarrow \{1,...,n\}$ and $\eta(i) \leq i$ for any $i$.
Denote by $(s)_{n-1} = (s_1,...,s_{n-1}) \in [0,1]^{\otimes n-1}$
a sequence of interpolating arguments and define corresponding
interpolating energies by the following inductive process
\beq
&{\cal E}&_\infty ((\lambda, \mu, \alpha)_n) \equiv {\cal E}_\infty
((\lambda, \mu, \alpha)_n),
\\ \nonumber
&{\cal E}&_{\infty,i}((\lambda, \mu, \alpha)_n) \equiv (1-s_i)
{\cal E}_\infty((\lambda, \mu, \alpha)_n \mid
(\lambda_i, \mu_i, \alpha_i)) + s_i{\cal E}_{\infty, i-1}
((\lambda, \mu, \alpha)_n)
\\ \nonumber
&{\cal E}&_{\infty, n-1} ((\lambda, \mu, \alpha)_n) \equiv
{\cal E}_\infty^n((s)_{n-1}).
\eeq
Define a function $f( \eta, (s)_{n-1})$ by
\beq
\left\{  \begin{array}{ll}
f(\eta, (s)_{n-1}) & = \prod_{i=2}^{n-1} s_{i-1} s_{i-2}...
s_{\eta(i)}, \,\,\,\mbox{for}\,\,\,n \geq 2\\
f(\eta, s_1) & = 1
\end{array} \right.
\eeq
where $s_{i-1} ...  s_{n(i)}$ is 1 if $\eta(i) = 1$.
The n-th virial coefficient $b_\infty^n$ can be written in the
following compact form
\beq
b_\infty^n(\beta) = \frac{1}{(-\beta)^{n-1} n}
\sum_{\eta \in T_{n}} \int d(s)_{n-1} \int
d_\infty^R(\lambda, \mu, \alpha)_N
\\ \nonumber
f(\eta, (s)_{n-1})) \prod_{i=1}^{n-1} {\cal E}(d(i+1))\mid
d(\eta(i))) \exp [-\frac{1}{\beta} {\cal E}_\infty^n((s)_{n-1})]
\eeq
where
\beq
\int d_\infty^R (\lambda, \mu, \alpha)_n ...
= \sum_{\mu_{1}\in 0_R} \int d\mu_{\mu_{1}}(\alpha_1)
\sum_{\begin{array}{c}
\lambda_2, \mu_2 \in \Omega_\infty\\
\mid \lambda_2 - \mu_2 \mid \leq R \end{array}} \int d
\mu_{\lambda_{2}-\mu_{1}} \ldots
\eeq
The following estimate is well known (see [24])
\beq
\sum_{\eta \in T_{n}} \int d(s)_{n-1} f(\eta, (s)_{n-1})
\leq e^{n-1}
\eeq
The induction steps defined in (3.46) involve the convex sums of
energies only, therefore the positive definiteness
of the intermediate and final energy function is preserved
in this process. This observation and estimate (3.50)
lead to the following bound on the n-th virial coefficient
\beq
\mid b_n^\infty \mid \leq \frac{1}{\mid \beta \mid^{n-1}}
\frac{e^{n-1}}{n} \parallel {\bm V}_\infty \parallel_1^{n-1}
\eeq
where
\beq
\parallel {\bm V}_\infty \parallel_1 \equiv \mid 0_R \mid
\sup_{\lambda \in 0_{R}}
(\int d \mid \mu_\lambda \mid (\alpha) \mid \alpha\mid)
\\ \nonumber
\cdot \sup_{\lambda \in 0_{R}} ( \sum_{\begin{array}{c}
\lambda', \mu' \in \Omega_{\infty}\\
\mid \lambda' - \mu' \mid \leq R \end{array}}
\int d \mid \mu_{\lambda' - \mu'}\mid (\alpha')\mid \alpha'\mid
\mid \tilde{{\bm V}}_\infty(0,\lambda \mid \lambda', \mu') \mid
\eeq
where
\beq
\tilde{{\bm V}}_\infty (\lambda, \mu \mid \lambda', \mu')
= {\bm C}_\infty (\lambda - \lambda') +
{\bm C}_\infty (\mu - \mu') - {\bm C}_\infty (\lambda - \mu')
- {\bm C}_\infty (\mu - \lambda').
\eeq
In this way we have proved:

\begin{proposisjon}
Assume hypotheses H2, H3 and H3$\gamma$ are satisfied and
moreover that $\parallel {\bm V}_\infty \parallel_1 < e^{-1}$,
then the virial expansion
\beq
P_\infty(\beta) = (-\beta) \cdot \sum_{n \geq 1} \frac{1}{n}
\sum_{\eta \in T_{n}} \int d(s)_{n-1}
\int d_\infty^R (\lambda, \mu, \alpha)^0_n
\\ \nonumber
f(\eta, (s)_{n-1}) \cdot \prod_{i=1}^{n-1} {\cal E}_\infty
(d(i+1) \mid d(\eta (i))
\exp [-\frac{1}{\beta} {\cal E}_\infty^n((s)_{n-1})]
\eeq
is absolutely convergent for any $\beta \in {\bm C}$ such that
$Re \beta > 0.$
\end{proposisjon}

{\bf Remark}\\
Define the Borel transform $\tilde{{\cal P}}_\beta(Z)$ generating
function for the virial coefficients by
\beq
\tilde{{\cal P}}_\beta(\xi) \equiv \sum_{n \geq 1}
\frac{(-\beta)^{n}\zeta^{n}b_{\infty}^{n}(\beta)}{n!}
\eeq
{}From the estimate (3.50) it follows that
$\tilde{{\cal P}}_\beta(\zeta)$ is an entire analytic function
on ${\bm C}$ in $\zeta$, for any
$\beta \in {\bm R}^+ \backslash \{0\}.$ Therefore applying the
inverse Borel transform to $\tilde{{\cal P}}_\beta$ we obtain
a function
${\cal P}_\beta(z) = (B^{-1}\tilde{{\cal P}}_\beta^{B^{-1}})(z)$
for which an expansion in powers of $z$
coincides with the virial expansion (3.54) for $z = 1$ and
moreover the expansion is a Borel summable as
$\beta \downarrow 0$:
\beq
{\cal P}_\beta(z)_{\mid z = 1} = \sum_{n\geq 1}^N
\frac{-(\beta)^{n}b_{\infty}^{n}(\beta)}{n!} + 0_N(\beta^{N+1})
\eeq
with $\mid 0_N(\beta^{n+1}) \mid \leq \mid C_N \mid \beta
\mid^{N+1}$ for some constant $C_N > 0 $.

{\bf Remark}\\
The Borel summability of the high-temperature expansion in
the classical statistical mechanics has been
studied in [26]. To prove Borel summability of the corresponding
low temperature expansion (3.54) one needs to expand
the corresponding Mayer kernels in powers of $\beta^{-1}$ and
then reformulate the whole expansion (3.54) in powers of
$\beta^{-1}$.

{\bf 3.3 The associated infinite volume Gibbs states.}\\
In this section we shall construct the infinite volume Gibbs
states describing dipole systems fulfilling all hypotheses
H0 - H3 above. It is well known that the
corresponding Gibbs states are determined by their correlation
functions, see e.g. [22]. Therefore
we concentrate on them in the following.

The finite volume correlation functions are given by:
\beq
\rho_N((\lambda, \mu, \alpha)_n) = (-\beta)^n {\bm E}_N^{\,\beta}
(\prod_{j=1}^n
\exp [i \beta^{-1/2}\alpha_j (y_{\lambda_{j}}- y_{\mu_{j}})]
\eeq
where the expectation denoted as ${\bm E}_N^\beta$ is defined as
\beq
{\bm E}_N^{\,\beta} (-) =
\frac{{\bm E}\,_N^0(- \exp [-\beta{\cal R}(\Omega_N\mid
\{\beta^{-1/2}y_{\lambda}\}_{\lambda \in \Omega_{N}}}
{{\bm E}\,_N^0 (\exp [- \beta{\cal R}(\Omega_N\mid
\{\beta^{-1/2}y_{\lambda}\}_{\lambda \in \Omega_{N}}}
\eeq

To be more general we consider also the possible influence of
the external dipole configurations on the thermodynamic limits
of the corresponding Gibbs states. Towards this goal, let us
consider a generic point
$\omega \in (\Omega_\infty  \times \Omega_\infty \times
{\bm R}\,^3)^\infty = {\cal D}_\rho(\infty)$,
where ${\cal D}_\rho(\infty)$ is defined as a configuration of
dipoles such that to every point
$(\lambda, \mu) \in \Omega_\infty \times \Omega_\infty$ there
is inserted at most $\rho$ dipoles with $\rho$ is some integer.
It follows from the finite length dipole approximation that for every
$\omega \in {\cal D}_\rho(\infty)$ there is only a finite number
of dipoles that can share the same end point
$\lambda \in \Omega_\infty$ and which contribute to the total
energy in a nontrivial way.

The restriction of a given dipole configuration
$\omega \in \Omega_\rho (\infty)$ to the set
$(\Omega_\infty \times \Omega_\infty \times {\bm R}\,^3)
\backslash \{\Omega_N \times \Omega_N \times {\bm R}\} \equiv
\Sigma^C_N$ will
be denoted by $\omega(N^c)$. The associated, finite volume,
conditional Gibbs
states are described by their conditional correlation functions
$\rho_N^\omega$ which are defined for $\omega \in \Omega_\rho(\infty)$
by
\beq
\rho_N^\omega((\alpha,\lambda,\mu)_n)=&(-\beta)^n
\cdot {\bm E}_\infty^0 \Bigl(\prod_{j=1}^n
\exp [i \beta^{- {1 \over 2}} \alpha_j
(y_{\lambda_j}- y_{\mu_j}) ]  \cr
&\cdot
\exp -\beta {\cal R}(\Omega_N\vert\cdot)
\cdot \exp -\beta {\cal F_N}(\omega(N^C))\bigr),  \cr
\eeq
where
\beq
{\cal F}_N(\omega(N_C)) \equiv
\sum_{(\lambda^{\prime}, \mu^{\prime}) \in {\tilde\Sigma}_N^C}
{\cal R}(\lambda^{\prime}-\mu^{\prime} \vert {y_{\lambda^{\prime}} -
y_{\mu^{\prime}} \over \sqrt\beta} )
\eeq
gives the energy of a given configuration of dipoles
$(\lambda, \mu, \alpha)_n$ inside $\Lambda_N$ with
an external configuration $\omega(N^c)$ realizing a particular
boundary condition. Due to the decay
properties of ${\bm V}_\infty$ and the assumption
$\omega \in {\cal D}_\rho(\infty)$, the
energy in (3.59) is finite.

The conditional partition function ${\bm Z}_N^\omega$ used in
(3.59) is given by the formula:
\beq
Z_N^{\omega} =
{ {\bm E}_{\infty}^0 \bigl( \exp -\beta {\cal R}(\Omega_N \vert
\cdot) \exp -\beta {\cal F}_N (\omega (N_C)) \bigr)
\over
{\bm E}_{\infty}^0 \bigl( \exp -\beta{\cal F}_N (\omega(N_C))
\bigr) }
\eeq
Performing the following, (complex) shift transformation
\beq
y_\lambda - y_\mu \rightarrow y_\lambda
- i {\bm C}_\infty(y_\lambda - y_{\lambda_{1}}) - y_\mu
- i {\bm C}_\infty (y_\mu - y_{\mu_{1}})
\eeq
in formula (3.59) and respectively
\beq
y_\lambda - y_\mu \rightarrow y_\lambda - i {\bm C}_N(y_\lambda
- y_{\lambda_{1}}) - y_\mu - i {\bm C}_N (\mu - \mu_{1})
\eeq
in the formula (3.57) and calculating the corresponding
Radon-Nikodym derivatives we obtain the following identities
\beq
\rho_N^\omega ((\alpha,\lambda,\mu)_n)
&= (-\beta)^n \exp
[-\beta^{-1} {\cal E}_{\infty}(d(1) \vert (\lambda,\mu,\alpha)_n
\cup \omega(N^C)) ]  \cr
&\cdot {\bm E}_N^0  \bigl( \prod_{j=2}^n e^{i \beta^{1 \over 2}
\alpha_j (y_{\lambda_j} - y_{\mu_j}) }
\exp - \beta {\cal R} (\Omega_N \vert e^{ - \beta^{-1}
{\cal E}_{\infty} (d(1) \vert \cdot)} -1 ) \bigr) \cr
& \exp - \beta {\cal F}_N (\omega (N^C) \vert e^{- \beta^{-1}
{\cal E}_{\infty} ( d(1) \vert \cdot) } -1 ),  \cr
\eeq
where
$$
{\cal F}_N(\omega(N_C)\vert e^{- \beta^{-1} {\cal E}_\infty
(d(1) \vert \cdot)}-1)
= \sum_{ {\lambda, \mu \in {\tilde\Sigma}_N^C }\atop {\vert
\lambda - \mu \vert \le R}} \int d\gamma
(\alpha_{\lambda-\mu}^\prime) e^{i \alpha_{\lambda- \mu}^{\prime}
( {y_{\lambda}^{\prime} - y_{\mu}^{\prime} \over \sqrt \beta} )}
\bigl( e^{ - \beta^{-1} {\cal E}_{\infty} (d(1) \vert
(\alpha_{\lambda - \mu}^{\prime}, \lambda, \mu)} -1 \bigr)
$$
and
\beq
\rho_N((\alpha, \lambda, \mu)_n)=&
(-\beta)^n \exp -\beta^{-1} {\cal E}_N (d(1) \vert (\alpha,
\lambda, \mu)_n)
\cr
& {\bm E}_0^{\beta} \bigl( \prod_{j=2}^n e^{i \beta^{{1 \over 2}}
\alpha_j ( y_{\lambda_j} - y_{\mu_j} ) }
\exp -\beta {\cal R} (\Omega_N \vert (e^{- \beta^{-1} {\cal E}
(d(1) \vert \cdot)} -1) \bigr)
\eeq
where
$$
{\cal R} (\Omega_N \vert e^{ - \beta^{-1} {\cal E}_N(d(1) \vert
\cdot )} -1)
\equiv
\sum_{ {\lambda,\mu \in \Omega_N} \atop {\vert \lambda - \mu
\vert \le R}} \int d\mu_{\lambda- \mu}(\alpha_{\lambda- \mu})
\cdot e^{i \alpha_{\lambda-\mu}^{\prime} ( {y_\lambda^\prime -
y_\mu^\prime \over \sqrt\beta} )}
( e^{- \beta^{-1} {\cal E}_N (d(1) \vert
(\alpha_{\lambda- \mu}^{\prime}, \lambda^{\prime},\mu)} -1 )
$$
in which, after an expansion, we recognize the well known
Kirkwood-Salzburg identities, see e.g. [22]. Using the method
of a dual pair of Banach spaces as explained in [25], a large
region of admissible $\beta$ will be determined in
which the rigorous comparison analysis of (3.64) with (3.65)
is possible.

For any $\xi > 0$ introduce a Banach space ${\bm E}_\xi$ consisting
of all sequences of functions
$(f_n((\lambda, \mu, \alpha)_n)_{n = 1,2,...})$ such that for any $n$,
any $\lambda, \mu \in \Omega_\infty$ the functions
$\tilde{f}_{\lambda, \mu} ((\alpha)_n) \equiv f_n((\lambda, \mu,
\alpha)_n)$ are measurable and such that the norm
$\parallel (f_{n})_{n} \parallel_\xi = \sup_n \xi^{-n} ess
\sup_{(\lambda,
\mu, \alpha)_{n}} \mid f_n((\lambda, \mu, \alpha)_n) \mid$
is finite. Several bounded linear operators will be defined to act
in the space ${\bm E}_\xi$. The infinite volume
Kirkwood-Salsburg operator $K_\infty(\beta)$ is defined by
\beq
(K_\infty((f_m)_m))_n ((\lambda, \mu, \alpha)_n) \equiv \exp
[-\frac{1}{\beta} {\cal E}_\infty ((\lambda_1, \mu_1, \alpha_1)]
\mid (\lambda, \mu, \alpha)_n) \\ \nonumber
\cdot \sum_{m \geq 0} \frac{1}{m!} \int d_\infty^R
(\lambda', \mu', \alpha')_m
\prod_{i=1}^m (\exp [-\frac{1}{\beta} {\cal E}_\infty
((\lambda_1, \mu_1, \alpha_1) \mid (\lambda_i', \mu_1',
\alpha_i'))]-1] \\ \nonumber
f_{n+m-1}((\lambda, \mu, \alpha)_{n-1}^*, (\lambda',\,\mu',
\alpha')_m)
\eeq
where
\beq
(\lambda, \mu, \alpha)_{n-1}^* \equiv ((\lambda_2, \mu_2, \alpha_2)
,...,(\lambda_n, \mu_n, \alpha_n)).
\eeq
The finite volume, unconditional Kirkwood-Salsburg operator $K_N$
is defined by:
\beq
(K_N((f_m)_m))_n ((\lambda, \mu, \alpha)_n) \equiv \exp
[-\frac{1}{\beta} {\cal E}_N ((\lambda_1, \mu_1, \alpha_1)
\mid (\lambda, \mu, \alpha)_n)]
\\ \nonumber
\cdot \sum_{m \geq 0} \frac{1}{m!} \int d_N^R (\lambda', \mu',
\alpha')_m \prod_{i=1}^m (\exp [-\frac{1}{\beta} {\cal E}_N
((\lambda_1, \mu_1, \alpha_1) \mid (\lambda_i', \mu_i',
\alpha_i'))]-1)
\\ \nonumber
f_{n+m-1}((\lambda, \mu, \alpha)_{n-1}^*, (\lambda',\,\mu', \alpha')_m).
\eeq
The energy factor operators are defined by:
\beq
E_N^\omega ((f_m)_m)_n ((\lambda, \mu, \alpha)_n) \equiv \chi_N
((\lambda, \mu)_n)\exp[-\frac{1}{\beta} {\cal E}_\infty
((\lambda_1, \mu_1, \alpha_1) \mid \omega (N^c))] \\
\cdot (E_N \cdot f)_n ((\lambda, \mu, \alpha)_n) \nonumber
\eeq
where
\beq
\chi_N (\lambda, \mu)_n) \equiv \left\{ \begin{array}{ll}
1 & (\lambda_i, \mu_i) \in \Omega_N \times \Omega_N
\mbox{ for all } i\\
0 & \mbox{ otherwise }. \end{array} \right.
\eeq
\beq
E_N^N((f_m)_m)_n ((\lambda, \mu, \alpha)_n) = \chi_N
((\lambda, \mu)_n) \exp [-\frac{1}{\beta}
{\cal E}_N((\lambda_1, \mu_1, \alpha_1) \mid (\lambda, \mu,
\alpha)_n)] \\ f_n((\lambda, \mu, \alpha)_n) \nonumber
\eeq
\beq
E_N((f_m)_m)_n ((\lambda, \mu, \alpha)_n) = \chi_N((\lambda, \mu)_n)
\exp [-\frac{1}{\beta}
{\cal E}_\infty((\lambda_1, \mu_1, \alpha_1) \mid (\lambda, \mu,
\alpha)_n)] \\
\cdot f_n((\lambda, \mu, \alpha)_n) \nonumber
\eeq
The projections $\prod_N$ are defined by
\beq
\Pi_N ((f_m)_m)_n((\lambda, \mu)_n) = \prod_{i=1}^{n} \chi_N
((\lambda_i, \mu_i))f_n((\lambda,\mu)_n)
\eeq
The finite volume conditional Kirkwood-Salsburg operator
$K_N^\omega$ is defined as follows:
\beq
K_N^\omega = E_N^\omega \cdot K_\infty \cdot \Pi_N .
\eeq
For further use we also define the operator $\tilde{K}_\infty$
through the relation
\beq
K_\infty = E_N \cdot \tilde{K}_\infty .
\eeq
Also the finite volume version of $\tilde{K}_\infty$ is defined by
\beq
K_N = E_N^N \cdot \tilde{K}_N.
\eeq
{}From the positive definiteness of ${\bm V}_\infty$ it follows that
we can split the set
$(\Omega_\infty \times \Omega_\infty \times {\bm R}\,^3)^{\otimes n}$
as $\cup_{j=1}^n \sum_j^n$, where
$\sum_j^n = \{(\lambda, \mu, \alpha)_n \in (\Omega_\infty \times
\Omega_\infty \times {\bm R}^3)^{\otimes n} \mid
\sum_{i\neq j}^n {\bm V}_\infty (\lambda_i \mu_i \alpha_i \mid
\lambda_j \mu_j \alpha_j) \geq
-2\sum_{i=1}^n \alpha_i^2 ({\bm C}_N(0) - {\bm C}_N(\lambda_i
- \mu_i))\}.$
Let then $\eta_j^n$ denote the characteristic function of $\sum_j^n$
and let $\theta_j^n = \eta_j^n/\sum_{j=1}^n \eta_j^n$. Let $S_k^n$
be defined on functions of $n$-varibles
$(\lambda, \mu, \alpha)_n$ as the circular permutations of
$k$-steps on the complex arguments $(\lambda, \mu, \alpha)$ of
these functions. Then the index juggling operator
$J_\infty$ is defined as
\beq
J_\infty ((f_m)_m)_n ((\lambda, \mu, \alpha)_n) =
\sum_{k=1}^n S_k^n((\theta_j^n((\lambda, \mu, \alpha)_n) f_n
((\lambda, \mu, \alpha)_n)
\eeq
In a similar way we define the Ruelle index juggling operator
${\cal J}_N$ which
selects a labelling of coordinates in such a way that
${\cal E}_N((\lambda_1, \mu_1, \alpha_1)\mid (\lambda_n, \mu_n,
\alpha_n)) \geq 0.$

Finally the following vectors are defined
\beq
\alpha_N^\omega = (\exp [- \frac{1}{\beta} {\cal E}_\infty
((\lambda_1, \mu_1, \alpha_1) \mid (\lambda, \mu, \alpha)_1
\cup \omega (N^c)),0,...)
\eeq
\beq
\alpha_N = (\exp [- \frac{1}{\beta} {\cal E}_N((\lambda_1,
\mu_1, \alpha_1) \mid (\lambda_1, \mu_1, \alpha_1))], 0,...,) .
\eeq
With this notation the equalities (3.64) and (3.65) can be
rewritten in the following way
\beq
((\rho_N^\omega)_n) = (-\beta) (\cdot \Pi_N \cdot
{\cal J}_\infty K_N^\omega \cdot \Pi_N)((\rho_N^\omega)_n)
+ (-\beta) \Pi_N \alpha_N^\omega
\eeq
and respectively
\beq
((\rho_N)_n) = -\beta (\Pi_N \cdot {\cal J}_N \cdot K_N \Pi_N)
((\rho_N)_n) - \beta \Pi_N \alpha_N .
\eeq
We expect that the infinite volume limits $((\rho_\infty)_n)$
of $((\rho_N)_n)$ should fulfill the following, infinite-volume
Kirkwood-Salsburg identities:
\beq
((\rho_\infty)_n) = -\beta J_\infty K_\infty ((\rho_\infty)_n)
- \beta \alpha_\infty^\phi
\eeq
and in fact we will show this for the infinite volume limits of
the conditional correlation functions, providing they exist.

The rigorous, comparison analysis of the identities (3.80), (3.81)
and (3.82), based on the use of methods of the dual pair of Banach
spaces as explained in [25] will be presented below. The
conclusions obtained in this way seem to be much richer then the
contraction map principle (see, i.e., [22])
ordinary used for such an analysis.

It is necessary to introduce the additional (technical) hypothesis
for making our analysis complete.\\[5 mm]
{\bf Hypothesis 4}\\
(H4) There exists a number $\xi_*$, possibly depending on $\rho$
and $R$ such that for any $\xi > \xi_*$ the family
$(\rho_N^\omega((\lambda, \mu, \alpha)_n)_n)_N$ is
weakly -* precompact in the space
${\bm E}_\xi$ for any $\omega \in D_ \rho(\infty)$.

Note that for the case $\omega = \phi$ and for the correlation
functions (3.57) H4 is fulfilled due to estimate (3.7) and the
Banach-Aloglou theorem. For a general $\omega \in D_\rho(\infty)$
the necessary estimates are still to be proved.

To formulate the main result of our analysis let us define
$\sigma_\xi(I_\infty K_\infty(\beta))$ the spectral set of the
operator $J_\infty K_\infty(\beta)$ in the corresponding
space ${\bm E}_\xi$ (see below) and let us define the set $H_\xi$
as:
\beq
H_\xi \equiv \{\beta \in {\bm C} \mid Re \beta > 0;\,\, \mid \beta
\mid < \xi\,\,\,\mbox{and}\,\,\, (-\beta)^{-1}
\notin \sigma_\xi(J_\infty K_\infty (\beta))\}
\eeq

\begin{teorem}
Let us assume that the hypotheses H0-H3, H3$\gamma$ and H4 are
all valid.

Then there exists a number $\xi'_*$ (possibly depending of $\rho$ and
$R$) such that for any
$\beta \in H_{\xi'_*}$ there exists a unique thermodynamic limit
$\rho_\infty^\omega$ of
$\rho_N^\omega((\lambda, \mu, \alpha)_n)$ which does not depend
in particular on $\omega \in  D_\rho (\infty)$
and is equal to the thermodynamic limit $\rho_\infty$ of
$(\rho_N((\lambda, \mu, \alpha)_n))$ which also exists
as a unique limit as $N \uparrow \infty$. The limits should
be understood as limits in the locally uniform, componentwise
topology of the space
${\bm E}_\xi$. The limiting correlation functions $\rho_\infty$
depend analytically on $\beta$,
provided $\beta \in H_{\xi_{*}^{'}}$, they are translationally
invariant and posses a cluster decomposition
property with respect to the translations of
$\Omega_\infty \times \Omega_\infty$.
\end{teorem}

Before passing	to the proof of this theorem let us note some
simple consequences which seem to be
relevant for the discussion of the problem of the convergence
of the linked cluster expansion worked out in this
section. For this goal let us define
\beq
C(\beta) \equiv \sup_{\lambda \in O_R} (\mid \sup \{\mid \alpha
\mid \, \mid \alpha \in \mbox{ supp }
d\mu_\lambda \}) \cdot \mid O_R \mid \mid
\\ \nonumber
(\int d_\infty^R \mid (\lambda ', \mu ', \mu ')\mid\,\mid
\exp [- \frac{1}{\beta} {\cal E}_\infty ((0, \lambda, \alpha)
\mid (\lambda ', \mu ', \alpha ')] -1 \mid)
\eeq
and
\beq
C\,'(\beta) \equiv \sup_N \sup_{\lambda \in O_R} (\mid \sup
\{\mid \alpha \mid \, \mid \alpha \in \mbox{ supp }
d\mu_\lambda \}) \cdot \mid O_R \mid
\\ \nonumber
(\int d_\infty^R \mid (\lambda ', \mu ', \mu ')\mid\,\mid
\exp [- \frac{1}{\beta} {\cal E}_N ((0, \lambda, \alpha)
\mid (\lambda ', \mu ', \alpha ')] -1 \mid).
\eeq
It follows from our hypothesis that these quantities
are finite (see the proof of Prop. 3.1).

Let us also define
\beq
H_\xi^C \equiv \{ \beta \in {\bm C} \mid Re\beta >
0\,\,\,\mbox{and}\,\,\,
\mid \beta \mid \leq \min \{\mid \xi \mid, C^{-1}(\beta)
e^{-1} < 1 \}\}
\eeq
and
\beq
H_\xi^{C'} \equiv \{ \beta \in {\bm C} \mid Re\beta >
0\,\,\,\mbox{and}\,\,\,
\mid \beta \mid \leq \min \{\mid \xi \mid, C'^{-1}(\beta)
e^{-1} < 1 \}\} .
\eeq

\begin{korollar}
Let us assume that all the assumptions of Theorem 3-3 are
fulfilled. Then $H_\xi^C\subset H_\xi(\beta)$ and
$H_\xi^{c'}\subset H_\xi(\beta)$, therefore all conclusions
of Theorem 3-3 are true for $\beta \in H_\xi^c$
(respectively $\beta \in H_\xi^{c'}$).
\end{korollar}

\begin{Proof}
It is very easy to check that for $\beta \in H_\xi^c$ the
corresponding Kirkwood-Salsburg operator
$-\beta I_\infty K_\infty$
is a contraction of the corresponding space ${\bm E}_\xi$ and
therefore $-\beta^{-1}\notin \sigma_\xi (J_\infty K_\infty)$.
The same argument applies for the corresponding
Kirkwood-Salsburg operator $J_NK_N$ as well.
\end{Proof}
Another corollary of Theorem 3-3 can be formulated as follows.
Let us define the conditional, finite-volume free energy density
$p_N^\omega$ as:
\beq
p_N^\omega = \frac{1}{\mid D_N(1)\mid} \ln {\bm Z}_N^\omega
\eeq
Then
\begin{korollar}
Assume that all hypotheses of Theorem 3-3 hold. Then for any
$\omega \in D_c(\infty)$ the unique thermodynamic limit
$p_\infty '(\beta) = \lim_{N\uparrow \infty} p_N^\omega (\beta)
= \lim_{N \uparrow \infty} p_N^\phi$ exists, provided
$\beta\in H_\xi$. Moreover the map
$H_\xi \ni \beta \rightarrow p_\infty (\beta)$ is analytic.
\end{korollar}

{\bf Proof of Theorem 3.3:}\\
The Banach space ${\bm E}_\xi$ is the dual of the Banach space
$^*{\bm E}_\xi$ which is formed
from all sequences  $(\psi_n((\lambda, \mu, \alpha)_n))$ of
functions defined on $(\Omega_\infty \times
\Omega_\infty \times {\bm R}\,^3)^{\otimes n}$
measurable in the charge coordinates $\alpha$ and such they have
a finite norm $^*\parallel\,\,\,\parallel_\xi$ given by:
\beq
^* \parallel (\psi_n)_n \parallel_\xi = \sum_{n \geq 0} \xi^n
\int d_\infty^R \mid (\lambda, \mu, \alpha) \mid\,\mid
\psi_n((\lambda, \mu, \alpha)_n) \mid.
\eeq
Define in the space ${\bm E}_\xi$ (for suitable $\xi$ to be
chosen later on) the following linear operators
\beq
(^*\tilde{K}_\infty((\psi_m)_m))_n((\lambda, \mu, \alpha)_n)
= \sum_{l = 0}^{n} \frac{1}{l!} \int
d_\infty^R (\lambda ', \mu ', \alpha ') \\ \nonumber
\prod_{i=1}^l [(\exp [-\frac{1}{\beta} {\cal E}_\infty ((\lambda_i,
\mu_i, \alpha_i) \mid (\lambda ', \mu ', \alpha '))] -1]
\\ \nonumber
\psi_{n+1-l}((\lambda ', \mu ', \alpha '), (\lambda_{l+1},
\mu_{l+1}, \alpha_{l+1}),... (\lambda_n, \mu_n, \alpha_n))
\eeq
and
\beq
^*\tilde{K}_N ((\psi_m)_m))_n((\lambda, \mu, \alpha)_n)
= \sum_{l = 0}^{n} \frac{1}{l!} \int d_N^R (\lambda ', \mu ',
\alpha ')
\\ \nonumber
\prod_{i=1}^l [\exp [(-\frac{1}{\beta} {\cal E}_N ((\lambda_i,
\mu_i, \alpha_i) \mid (\lambda ', \mu ', \alpha '))] -1]
\\ \nonumber
\psi_{n-1-l}((\lambda ', \mu ', \alpha '), (\lambda_{l+1},
\mu_{l+1}, \alpha_{l+1}),... (\lambda_n, \mu_n, \alpha_n)) .
\eeq

Calculating the corresponding dual operators to the operators
$^*\tilde{K}_\infty$ and $^*\tilde{K}_N$ we arrive at the following
equalities
\beq
(^*\tilde{K}_\infty)^* = \tilde{K}_\infty\,\,\,\mbox{and}\,\,\,
(^*\tilde{K}_N)^* = \tilde{K}_N.
\eeq
{}From the estimates
\beq
\parallel \tilde{K}_\infty \parallel_\xi \leq \xi^{-1} \exp
[\xi {\bm C}(\beta)]
\eeq
\beq
\parallel \tilde{K}_N \parallel_\xi \leq \xi^{-1} \exp [\xi
{\bm C}(\beta)]
\eeq
the continuity of operators $^*\tilde{K}_\infty$ and
$^*\tilde{K}_N$ in the space $^*{\bm E}_\xi$ and for any
$\xi > 0 $ follows.

Let us choose a value of $\xi$ such that the family
$(\rho_N^\omega)_N \subset {\bm E}_\xi$ is
weakly -$^*$ precompact and let further $\rho_\infty^\omega$ be
any of the accumulation points.
As a result of the fact that finite-component sequences of
compactly supported functions in the space
$^*{\bm E}_\xi$ form a dense subspace and the local decay
properties of ${\cal E}_\infty$ (respectively
${\cal E}_N)$ we easily see that:
\beq
{^*\Pi_N} \cdot {^*K_\infty}\cdot {^*E_N^\omega} \cdot {^*J_N}
\cdot {^* \Pi_N} \longrightarrow_{\!\!\!\!\!\!\!\!\!{\mbox{strongly}}}
{^*K_\infty}\cdot {^*J_\infty}
\eeq
and
\beq
{^*\Pi_N}\cdot{^*K_N} \cdot {^*J_N} \cdot {^*\Pi_N}
\stackrel{\longrightarrow}{\mbox{strongly}}
{^*K_\infty}\cdot{^*J_\infty}
\eeq
as $N \uparrow \infty$ and where $^*\prod _N,\, ^*E_N^\omega$ and
$^*I_\infty,\,^*I_N$
are the corresponding predual of the operators
$\prod_N,\, E_N^\omega$ and $J_\infty^*,\, J_N$ whose
existence and form of action on the space $^*{\bm E}_\xi$, can be
easily determined from the
very definitions  (3.73), (3.69),(3.77) and remark after (3.77).

Taking into account that
$\alpha_N^\omega \rightarrow \alpha_\infty^\phi$ in
the weak -$^*$ topology of ${\bm E}_\xi$ and the convergence (3.92)
we conclude that any $\rho_\infty^\omega$ must fulfill the identity:
\beq
\rho_\infty^\omega = (-\beta) I_\infty K_\infty \rho_\infty^\omega
- \beta \alpha_\infty^\phi .
\eeq
Therefore assuming that $\beta \in H_\xi (\beta)$ we conclude
for the existence of a unique thermodynamic limit
$\rho_\infty^\omega$ which does not depends of $\omega$, provided
$\omega \in {\cal D}_\rho(\infty)$ and moreover
$\rho_N^\omega \rightarrow \rho_\infty^\omega$ in the weak -$^*$
topology. The same reasoning applies to $\rho_N$ as well
with the conclusion that the corresponding thermodynamic limit
$\rho_\infty = \rho_\infty^\omega$. Standard application of the
Mayer-Montroll identities (see i.e. [22], [26]) then improves
the proven weak -$^*$ convergence
$\rho_N^\omega \rightarrow \rho_\infty^\omega$ to the
locally uniform componentwise as stated in Theorem. The remaining
assertions are easier and will not be proven here.

Effective applications of Theorem 3-3 amounts to the study of the
spectral properties of the operator
$J_\infty K_\infty$ in the space ${\bm E}_\xi$. Using the ideas and
methods of [27, 28] it can be proved that
$\sigma_\xi (\prod (\Lambda) J_N K_N^\phi \prod (\lambda) \equiv
\{ - \beta \in {\bm C} \mid Z_N^{fl}((- \beta)^{-1}) = 0\}$.
However the localisation and the flow of zeros of the partition
function $Z_N^{fl}$ as
$N \uparrow \infty$ is outside the scope of the present paper.
It is not known to the authors whether the region
$H(\beta)$ is close to the low temperature region.

It should be mentioned that in the physical literature the limit
temperature $T \uparrow \infty$
and $Z \uparrow \infty$ has been studied before and is known as
Vlasov limit in theory of
plasma physics [29]. The Vlasov limiting region exactly
corresponds to the low temperature behaviour of dipole systems
studied by us. Whether
one can extract from the Vlasow theory some relevant information
to the dipole systems describing fluctuation from the crystalline
ground states remains to be studied.

\chapter{Concluding remarks}
In this section we shall clarify the meaning of the hypotheses
that we stated in the course of our analysis and put our work
in a certain perspective from which further possible developments
and the meaning of the low temperature expansion introduced here
become clearer.

First we shall discuss hypothesis $H3$.
The Taylor remainder (2-41) can be written as follws
\beq
{\cal R}_{\Omega_{N}}(\lambda - \mu \mid \frac{y_{\lambda}-
y_{\mu}}{\sqrt{\beta}}) = \int_0^1dt \frac{(1-t^{2})}{2!} [D^3V]
(\lambda + ty_\lambda - \mu -ty_\mu)\,(y_\lambda - y_\mu)^{3}
\eeq
Due to the geometry of ${\bm R}\,^3$,
even if $V$ has compact support the size of the fluctuations
$y_\lambda$ can be arbitrary. The only
restriction is that
$\lambda + y_\lambda - \mu - y_\mu \in \mbox{ supp } V$.
This leads to the question whether the Taylor remainder (4.1)
can be a bounded function in the variable $y_\lambda - y_\mu$.
{}From the elementary estimate
\beqq
\mid \lambda + y_\lambda - \mu - y_\mu \mid \geq \mid
\mid\!\lambda - \mu\!\mid - \mid\!y_\lambda - y_\mu\!\mid \mid
\eeqq
and the assumption that $V$ is central it follows that this can
be realized. As an example let us take:
\beq
f_\varepsilon (x) = \left \{ \begin{array}{l}
\exp [- \frac{1}{-\mid x \mid^2 + \varepsilon^{2}}]
\,\,\,\mbox{for}\,\,\,\mid x \mid < \varepsilon.\\
0\,\,\,\mid x \mid \geq \varepsilon \end{array} \right.
\eeq
Then from (4.2) we obtain
\beqq
\mid f_\varepsilon (\lambda - \mu + y_\lambda - y_\mu)\mid \leq
\exp [- \frac{1}{\varepsilon^{2} - \mid\mid\!\lambda - \mu\!\mid
- \mid\!y_\lambda - y_\mu\!\mid \mid^{2}}]
\eeqq
if $\mid \lambda - \mu + y_\lambda - y_\mu \mid \geq \varepsilon$
and $f_\varepsilon (\lambda - \mu + y_\lambda - y_\mu ) = 0$ if
$\mid \lambda - \mu + y_\lambda - y_\mu \mid \geq \varepsilon.$
This forces an exponentially fast decay of
$\mid f_\varepsilon (\lambda-\mu + y_\lambda - y_\mu) \mid\,\mid
(y_\lambda - y_\mu)^3\mid$ as
$\mid y_\lambda - y_\mu \mid \uparrow \infty$ provided
$\lambda, \mu$ are kept fixed.\\
The partition function (2.30) can be rewritten as:
\beq
{\bm Z}^{fl}_N (\beta) = {\bm Z}^{bd}_N (\beta)
\frac{{\bm Z}^{fl}_N (\beta)}{{\bm Z}^d_N (\beta)}
\frac{{\bm Z}^{d}_N (\beta)}{{\bm Z}^{bd}_N (\beta)}
\eeq
therefore the corresponding free energy density is
\beq
p^{fl}_N (\beta) = \frac{1}{\beta \mid \Lambda_N \mid}
ln {\bm Z}^{bd}_N + \frac{1}{\beta \mid \Lambda_N\mid} ln
\frac{{\bm Z}^{fl}_N (\beta)}{{\bm Z}^d_N (\beta)}
\times \frac{1}{\beta \mid \Lambda_N\mid} ln
\frac{{\bm Z}^{d}_N (\beta)}{{\bm Z}^{bd}_N (\beta)}
\eeq
The results of section 3 enables us to control rigorously
\beq
\lim_{N \rightarrow \infty}
\frac{1}{\beta \mid \Lambda_N \mid} ln {\bm Z}^{bd}_N
(\beta ) \equiv p^{sfl}_\infty (\beta)
\eeq
which we call small fluctuations contribution to
$\lim_{N \to \infty} p^{fl}_N (\beta) \equiv p^{fl}_\infty
(\beta)$ provided it exists. It is reasonable to expect that
\beq
\lim_{N \rightarrow \infty} \frac{1}{\mid \Lambda_N \mid}
ln \frac{{\bm Z}^{fl}_N (\beta)}{{\bm Z}^d_N (\beta)} = 0
\eeq
therefore the only (but highly nontrivial) problem to solve is to
show (with some additional hypothesis on $[D^2V]$) the existence
of the limit
\beq
\lim_{N \rightarrow \infty} \frac{1}{\mid \Lambda_N \mid}
ln \frac{{\bm Z}^d_N (\beta)}{{\bm Z}^{bd}_N (\beta)}
\equiv p^{lfl}_\infty (\beta)
\eeq
This limit be called large fluctuations free energy density. The
superstability like hypothesis on $[D^2V]$ can produce a
probability like estimates that make the appearance of very
large dipoles very unprobable and this might be sufficient for
the proof of the existence of $p^{lfl}_\infty (\beta)$.

Let $\mu^\beta$ be the corresponding canonical Gibbs ensemble
measure for the corresponding real system of particles at inverse
temperature $\beta$ and with the density $\rho$. Assume that the
lattice ${\cal L}$ is the ground state configuration. We
will say that the local finite configuration $\omega$ is an
almost ${\cal L}$--crystalline configuration iff
for sufficiently $N_\omega > 0$ a map
$$
l^\omega_{\cal L} : \omega_{\uparrow R^3 \backslash [- N_\omega ,
N_\omega]} \rightarrow {\cal L} \backslash \omega_N
$$
(where $\omega_{\uparrow \Lambda}$ is the restriction of
$\omega$ to the set $\Lambda$) defined by:
$$
l^\omega_{\cal L} (x_i) = l^\omega (i)
$$

where:
$l^\omega (i) = \lambda$ if there exists only one
$\l \in {\cal L}$ such that
$\mid x_i - \lambda \mid = \min \{ \mid x_i - \lambda \mid
\mid \lambda ' \in {\cal L} \}$ and if there are several
such $\lambda \in {\cal L}$ on which the minimum of
$\{ \mid x_i - \lambda ' \mid \mid \lambda ' \in {\cal L} \}$
is achieved the $\lambda$ is choosen to be the first of them in
the natural lexicographic order given to ${\cal L}$.
\begin{center}
\underline{exists} and is \underline{bijective}.
\end{center}
A Borel subset $\Xi$ of the space of all locally finite
configurations is called uniformly ${\cal L}$--crystalline iff
there exists
$N(\Xi) \equiv \sup_{\omega \in \Xi} N_\omega < \infty$.
If we suppose that for sufficiently small $\frac{1}{\beta}$ the
Gibbs measure $\mu^\beta$ is supported by some uniformly
crystalline subset then of course the problem of large
fluctuations will be overcomed by a suitable relabeling procedura.
However at present only a very limited knowledge on the support
properties of the corresponding Gibbs measures is available,
which is not sufficient for solving this problem.

Now we turn to the discussion of H1. In [32,33] Katz and Duneau
have showed that the set of smooth central potentials with compact
support that leads to an energy function ${\cal E}((x)_n)$
possessing Morse function properties is residual in the space of
$C^\infty({\bm R}^+, {\bm R})$ functions equipped with the
Whitney topology. Moreover they showed that the Morse character of
${\cal E}((x)_n)$ is precisely the necessary and
sufficient condition for having variations of potentials which gives
rise to a continuous trajectory of equlibria, preserving
the crystalline symmetries of the starting equilibrium
configurations. The abstract results of [32,33] substantiate
our motivation for imposing Hypothesis 1 onto our scheme.

That the set of potentials for which H2 holds is large is
expressed in the following proposition.

\begin{proposisjon}
The set of potentials  $\Phi \in C^\infty({\bm R}^+, {\bm R})$
for which H2 is true, is an nonvoid open set in the Whitney topology.
\end{proposisjon}
It would of course be interesting to produce explicit examples
satisfying all assumptions we made, in particular the uniformity
in N involved in H1, H2.

Some preliminary results on this problem are contained in [21]
and a more complete discussion of this problem is now in
preparation [34].

Another important question concerns the stability of the
crystalline symmetry in the low temperature region in the
following sense. In the Pirogov-Sinai theory [35] powerful
probabilistic arguments are worked out to show
that the corresponding pure phases do not fluctuate much around
the ground state configuration in the low temperature region
(This is expressed as follows: The probability of the occurrence
a large contour $\Gamma$ with length $\mid\!\Gamma\!\mid$ is
less then $\exp (- \beta \mid\!\Gamma\! \mid)$.
It is an important problem to develop systematically methods which
show that the typical configurations of a
real system (in the thermodynamic limit) in low temperature thermal
equilibrium do not fluctuate much around the
corresponding crystalline ground state configuration. But an even
more important and harder problem is to introduce an equivalent
notion of the Peierls stability condition of `small' perturbations
of the ground state which leads to stability of crystalline symmetry
in the low temperature region for the corresponding Gibbs state.

\newpage

{\bf Acknoledgements}\\
Raphael  Hoegh--Krohn  discussed  his  ideas   on  the	subject  with  several
mathematicians	and physicists,  in particular	with R.L.  Dobrushin and  S.B.
Shlosman, during a stay in moscow (with the first author) in the fall of 1987.
\medskip
We  are also  greatful to  N. Risebro  for providing  us with  some notes of a
course given by  Raphael in Oslo in 1986--87.  Stimulating discussions with Z.
Haba, E. Lieb, E. Seiler and B. Zegarlinski are also gratefully acknowledged.
\medskip
The financial support of the German--Polish Exchange Programm (Karlsruhe), KBN
Grant  Nr.  200  60  9101,  of	the  BiBoS Research Center (Bielefeld/Bochum),
Cerfirm     (Locarno),	   EC	  Mobility     Grant	 7022,	   SFB	   237
(Essen--Bochum--D\"usseldorf), Centre de  Physique Theorique--CNRS,
Universit\'e d'Aix--Marseille II  and Universit\'e de Provence  (Marseille),
NAVF (Norway),
Graduierten Kolleg  (DFG, Bochum) and  Swedish Research Council  is gratefully
acknowledged.

{\bf References:}
\begin{description}
\item[[1]] {\bf Anderson, P.W.,} Basic Notions of Condensed Matter Physics,
Benjamin 1984.\\
\item[[2]] {\bf Simon, B.,} Fifteen problems in Mathematical physics. In:
Perspectives in Mathematics. Anniversary of
Oberwolfach, 1984, (eds. W. Jager, J. Moserr, R. Remment), Birkh\"{a}user 1984,
pp. 442.\\
\item[[3]] {\bf Ventevogel, W.J.,} On the configuration of a one-dimensional
system of interacting particles
with minimum potential energy per particle. Physica {\bf 92A} (1978) 343-361.\\
\item[[4]] {\bf Ventevogel, W.J., Nijboer B.R.A.} On the configuration of
systems of interacting particles with minimum
potential energy per particle. Physica {\bf 98A} (1979) 274-288.\\
\item[[5]] {\bf Ventevogel, W.J., Nijboer B.R.A.}
On the configuration of systems of interacting particles with minimum potential
energy per particle.
Physica {\bf 99A} (1979) 569-580.\\
\item[[6]] {\bf Gardner, C.S., Radin, C.,} The infinite-volume ground state
of the Lennard-Jones potential. J. Stat. Phys. {\bf 20} (1979) 719-724.\\
\item[[7]] {\bf Hamrick, G.C., Radin, C.,} The symmetry of ground states under
perturbations. J. Stat. Phys.
{\bf 21} (1979) 601-607.\\
\item[[8]] {\bf Heitmann, R.C., Radin, C.,} The ground state for sticky disks.
J. Stat. Phys. {\bf 22} (1980) 281-287.\\
\item[[9]] {\bf Radin, C.,} The ground state for soft disks. J. Stat. Phys.
{\bf 26} (1981) 365-373.\\
\item[[10]] {\bf Wagner, H-J.,} Crystallinizy in two dimensions: a note on
paper by C. Radin. J. Stat. Phys. {\bf 33} (1983) 523-526.\\
\item[[11]] {\bf Radin, C.,} Crystals and quasicrystals. A continuum model.
Comm. Math. Phys. {\bf 105} (1986) 385-390.\\
\item[[12]] {\bf Burkov, S.E.,} One dimensional model of the quasicrystalline
alloy. J. Stat. Phys. {\bf 47} (1987) 409-438.\\
\item[[13]] {\bf Kennedy, T., Lieb, E.,} An itinerant electron model with
crystalline or magnetic long range order.
Physica {\bf 138A} (1986) 320-358.\\
\item[[14]] {\bf Lieb, E.,} A model for crystallization. A variation on the
Hubbard model. In: Proceeding of the
VIII'th International Congress on Mathematical Physics, Marseille 1986, (eds.
M. Mebkhout, R. S\'{e}n\'{e}or), World Scientific,
Singapore 1987, pp. 185 - 196.\\
\item[[15]] {\bf Radin, C.,} Low temperatures and the origin of crystalline
symmetry, Int. J. Mod. Physics B {\bf 1} (1987) 1157 - 1191.\\
\item[[16]] {\bf Fr\"{o}hlich, J., Spencer, T.,} On the statistical mechanics
of classical Coulomb and dipole gases,
J. Stat. Phys. {\bf 24} (1981) 617 - 701.\\
\item[[17]] {\bf Fr\"{o}hlich, J., Spencer, T.,}
The Kosterlitz-Thouless transition in two-dimensional Abelian spin systems and
the Coulomb gas, Comm.
Math. Phys. {\bf 81} (1981) 527 - 602.\\
\item[[18]] {\bf Park, Y.M.,} Lack of screening in the continuous dipole
systems, Comm. Math. Phys. {\bf 70} (1979) 161-167.\\
\item[[19]] {\bf Brydges, D.C., Federbush, P.,}
Debye screening. Comm. Math. Phys. {\bf 73}
(1980) 197 - 246.\\
\item[[20]] {\bf Gawedzki, K., Kupiainen, A.,} A rigorous block spin approach
to massless lattice theories. Comm. Math. Phys. {\bf 77}
(1980) 31 - 64.\\
\item[[21]] {\bf Albeverio, S., H\o egh-Krohn, R., Holden, H., Kolsrud, T.,
Mebkhout, M.,}
A remark on the formation
of crystals at zero temperature in: Stochastic Methods in Mathematics and
Physics.
Proceedings of the XXIV Karpacz Winter School
(eds. R. Gielerak, W. Karwowski), World Scientific, Singapore
(1989), pp. 211-229.\\
\item[[22]] {\bf Ruelle, D.,} Statistical Mechanics. Rigorous Results.
Benjamin, New York 1969.\\
\item[[23]] {\bf Albeverio, S., H\o egh-Krohn, R.,}
Uniqueness of the physical vaccum and the Wightman functions in the infinite
volume limit for some
nonpolynomial interactions, Comm. Math. Phys. {\bf 30} (1973) 171-200.\\
\item[[24]] {\bf Brydges, D., Federbush, P.,}
A new form of the Mayer expansion in classical statistical mechanics. J. Math.
Phys. {\bf 19} (1978) 2064 - 2067.\\
\item[[25]] {\bf Gielerak, R.,} Uniqueness theorem for a class of continuous
systems, Physica A {\bf 189} (1992), 348--366.
\item[[26]] {\bf Wagner,} Borel Summability of the High Temperature Expansion
for Classical Continous Systems, Comm. Math. Phys.{\bf82}
(1981)183 \\
\item[[27]] {\bf Pastur, L.A.,} Theor. Math. Phys. {\bf 18} (1974) 233-242.\\
\item[[28]] {\bf Zagrebnov, V.A.,} Spectral Properties of
Ruelle-Kirkwood-Salsburg and Kirkwood-Salsburg operator. J. Stat. Phys. {\bf
27} (1982) 577-591.\\
\item[[29]] {\bf Nazin, G.I., Njashin, A.F.,}
The Bogolubov equation and the Vlasov equation in equilibrium statistical
physics. Rep. Math. Phys. {\bf 21}
(1985) 79-89.\\
\item[[30]] {\bf Ruelle, D.,} Superstable interactions in classical statistical
mechanics. Comm. Math. Phys. {\bf 18}
(1970) 127-159.\\
\item[[31]] {\bf Lebowitz, J.L., Presutti, E.,} Statistical Mechanics of
Unbounded Continous Spin Systems. Comm. Math. Phys. {\bf 78} (1970) 151.\\
\item[[32]] {\bf Duneau, M., Katz, A.,}
Generic properties of classical n-body systems in one dimension and crystal
theory, Ann. Inst. H. Poincar\'{e} Sect. A.
{\bf 37} (1982) 249 - 270.\\
\item[[33]] {\bf Katz, A., Duneau, M.,}
Stability of symmetries for equilibrium configurations of N particles in three
dimensions. J. Stat. Phys. {\bf 29} (1982)
475 - 498.\\
\item[[34]] {\bf Albeverio, S., Gielerak, R., Holden, H.,} in preparation.\\
\item[[35]] {\bf Sinai Ya.,} Theory of Phase Transitions: Rigorous Results,
Pergamon, New York 1982.\\
\end{description}

\end{document}